\documentclass[conference,compsoc]{IEEEtran}
\IEEEoverridecommandlockouts
\usepackage{cite}
\usepackage{amsmath,amssymb,amsfonts}
\usepackage{algorithmic}
\usepackage[]{algorithm2e}
\usepackage{graphicx}
\usepackage{textcomp}
\usepackage{xcolor}
\usepackage{todonotes}
\usepackage{booktabs}
\usepackage{ntheorem}
\usepackage{subfigure}
\usepackage{blindtext}
\usepackage{csquotes}
\usepackage[normalem]{ulem}
\usepackage{multirow}
\usepackage{microtype}
\usepackage{balance}

\theoremstyle{definition}
\newtheorem{definition}{Definition}

\newtheorem{observation}{Observation}

\def\BibTeX{{\rm B\kern-.05em{\sc i\kern-.025em b}\kern-.08em
    T\kern-.1667em\lower.7ex\hbox{E}\kern-.125emX}}
    
\DeclareMathOperator*{\argmax}{arg\,max}
\DeclareMathOperator*{\argmin}{arg\,min}

\newcounter{myctr}
\newenvironment{mylist}{\begin{list}{(\textbf{\alph{myctr}})}
{\usecounter{myctr}
\setlength{\topsep}{1mm}\setlength{\itemsep}{0.5mm}
\setlength{\parsep}{0.5mm}
\setlength{\itemindent}{1mm}\setlength{\partopsep}{0mm}
\setlength{\labelwidth}{0mm}
\setlength{\leftmargin}{1mm}}}{\end{list}}

\begin{document}
\title{Message Time of Arrival Codes: \\ A Fundamental Primitive for Secure Distance Measurement}
% Not sure if we should call this codes. Codes produce a sequence of symbols based on a number of bits. Hypothesis test is then applied to each symbol. We can't do this. I see it more as a spreading sequence applied to the pulse.

%\author{\IEEEauthorblockN{Patrick Leu}
%\IEEEauthorblockA{Department of Computer Science\\
%ETH Zurich}
%\and
%\IEEEauthorblockN{Mridula Singh}
%\IEEEauthorblockA{Department of Computer Science\\
%ETH Zurich}
%\and
%\IEEEauthorblockN{Marc Roeschlin}
%\IEEEauthorblockA{Department of Computer Science\\
%ETH Zurich}
%\and
%\IEEEauthorblockN{Kenneth G. Paterson}
%\IEEEauthorblockA{Department of Computer Science\\
%ETH Zurich}
%\and
%\IEEEauthorblockN{Srdjan Capkun}
%\IEEEauthorblockA{Department of Computer Science\\
%ETH Zurich}}


\author{\IEEEauthorblockN{Patrick Leu,
Mridula Singh,
Marc Roeschlin, 
Kenneth G. Paterson, and
Srdjan \v{C}apkun}
\IEEEauthorblockA{Department of Computer Science\\
ETH Zurich}
%Telephone: (800) 555--1212, Fax: (888) 555--1212}
%\IEEEauthorblockA{\IEEEauthorrefmark{4}Tyrell Inc., 123 Replicant Street, Los Angeles, California 90210--4321}
}


\maketitle

\begin{abstract}
Secure distance measurement and therefore secure Time-of-Arrival (ToA) measurement is critical for applications such as contactless payments, passive-keyless entry and start systems, and navigation systems. 
This paper initiates the study of Message Time of Arrival Codes (MTACs) and their security. 
MTACs represent a core primitive in the construction of systems for secure ToA measurement. 
By surfacing MTACs in this way, we are able for the first time to formally define the security requirements of physical-layer measures that protect ToA measurement systems against attacks. 
Our viewpoint also enables us to provide a unified presentation of existing MTACs (such as those proposed in distance-bounding protocols and in a secure distance measurement standard) and to propose basic principles for protecting ToA measurement systems against attacks that remain unaddressed by existing mechanisms. We also use our perspective to systematically explore the tradeoffs between security and performance that apply to all signal modulation techniques enabling ToA measurements.
\end{abstract}


% \fi
\IEEEpeerreviewmaketitle

% Mtac name, #1: MTAC type (e.g., A)
\newcommand{\mtac}[1]{MTAC-#1}

% Forge experiment, #1 MTAC type (e.g., A), #2: Set of algorithms (e.g., \Pi), #3: Forger (e.g., A(\alpha))
\newcommand{\mtacforge}[3]{\mtacalgname\mathit{_{#1}-forge}(#2,#3)}

% Forge experiment, #1: MTAC type (e.g., A), #2: Forger (e.g., A(\alpha)), #3: Advancment goal (e.g., \delta), #4: Set of algorithms (e.g., \Pi), #5: Security parameter (e.g., n_p)
\newcommand{\mtacforgealt}[5]{\mtacalgname\mathit{_{#1}-forge}_{#2,#3,#4}(#5)}

% Forge experiment, KP: 
\newcommand{\mtacforgeKP}[4]{\mtacalgname\mathit{\text{-}{#1}\text{-}forge}_{#2,#3}(#4)}

% Advantage
\newcommand{\adv}{Adv}

% Forger advantage, #1: MTAC (e.g, MTAC-A), #2: Advancement goal (e.g., \delta) #3: Set of algorithms (e.g., \Pi), #4: Forger (e.g., A(\alpha)), #5: Security parameter length (e.g., n_p)
\newcommand{\mtacadv}[5]{\mathit{\adv}^{#1}_{#2,#3}(#4,#5)}

% Forger advantage, KP 
\newcommand{\mtacadvKP}[4]{\mathit{\adv}^{#1}_{#2,#3}(#4)}

% Max forger advantage (insecurity function), #1: MTAC (e.g., MTAC-A), #2: Set of algorithms (e.g., \Pi), #3: Number of queries, #4: Runtime, #5: Security parameter length (e.g., n_p)
\newcommand{\mtacadvmax}[5]{\textbf{\adv}^{#1}_{#2}(#3,#4,#5)}

% Max forger advantage (insecurity function), KP
\newcommand{\mtacadvmaxKP}[7]{\textbf{\adv}^{#1}_{#2}(#3,#4,#5,#6,#7)}

% Max forger advantage (insecurity function), KP -- Delay Attack
\newcommand{\mtacdelmaxKP}[6]{\textbf{\adv}^{#1}_{#2}(#3,#4,#5,#6)}

% Forger name
\newcommand{\frgrname}{A}

% Forger, #1: Observation delay (e.g., \alpha)
\newcommand{\frgr}[1]{\frgrname(#1)}

% Transmit signal
\newcommand{\txsig}{\mathbf{c}}

% Transmit signal sample
\newcommand{\txsigs}[1]{c_{#1}}

% Receive signal
\newcommand{\rxsig}{\txsig '}

% Receive signal sample
\newcommand{\rxsigs}[1]{\txsigs{#1}'}

% Superposition signal
\newcommand{\spsig}{\txsig ''}

% Superposition signal sample
\newcommand{\spsigs}[1]{\txsigs{#1}''}

% Set of queries
\newcommand{\qryset}{Q}

% Forger with arg, #1: Observation delay (e.g., \alpha)
\newcommand{\frgrarg}[1]{\frgr(#1)}

% Set of algorithms
\newcommand{\algset}{\Pi}

% Generation algorithm name
\newcommand{\genalgname}{\mathit{Gen}}

% Mtac algorithm name
\newcommand{\mtacalgname}{\mathit{Mtac}}

% Verification algorithm name
\newcommand{\vrfyalgname}{\mathit{Vrfy}}

% Verification algorithm, #1: Key (e.g., k), #2: Message (e.g., m), #3 Forger signal (e.g., \mathbf{c}')
\newcommand{\vrfyalg}[3]{{\vrfyalgname}_{#1}(#2,#3)}

% Advancement mode ID
\newcommand{\advid}{A}

% Delay mode ID
\newcommand{\delid}{D}

% Observation delay
\newcommand{\obsdel}{\alpha}

% Advancement goal
\newcommand{\advgoal}{\delta}

% Security parameter length
\newcommand{\secprmlen}{n}

% Frame length
\newcommand{\frmlen}{n_p}

% Security parameter
\newcommand{\secprm}{1^{\secprmlen}}

% Key
\newcommand{\key}{k}

% Transmit message
\newcommand{\txmsg}{m}

% Receive message
\newcommand{\rxmsg}{\txmsg '}

% Transmit bit, #1: Index
\newcommand{\txbit}[1]{b_{#1}}

% Receive bit, #1: Index
\newcommand{\rxbit}[1]{\txbit{#1} '}

% Number of bits in message
\newcommand{\nbit}{n_b}

% Verification output
\newcommand{\vrfybit}{b}

% Residual stochastic process
\newcommand{\ressprc}{\mathbf{R}}

% Noise stochastic process
\newcommand{\nsprc}{\mathbf{N}}

% Forger error stochastic process
\newcommand{\frgrsprc}{\mathbf{A}}

% Residual signal
\newcommand{\ressig}{\mathbf{r}}

% Noise stochastic process
\newcommand{\nprc}{\mathcal{N}}

% Forger error stochastic process
\newcommand{\frgrprc}{\mathcal{A}}

% Number of pulses per bit
\newcommand{\nppb}{n_{ppb}}

% Symbol, #1: Bit value
\newcommand{\symb}[1]{\mathbf{s}_{#1}}

% Symbol frame
\newcommand{\symbfrm}{\mathbf{b}}

% XOR sequence
\newcommand{\xorseq}{\mathbf{x}}

% Rx-side signal template
\newcommand{\rxtmpl}{\hat{\txsig}}

% Performance region
\newcommand{\perfreg}{\mathcal{P}}

% Performance level
\newcommand{\perflvl}{p}

% Distortion
\newcommand{\dstrt}{\mathcal{D}}

% Threshold
\newcommand{\thrs}[1]{T_{#1}}

% Effective threshold
\newcommand{\thrseff}[1]{\hat{T}_{#1}}

% Prover
\newcommand{\prvr}{P}

% Verifier
\newcommand{\vrfr}{V}

% Transmitter
\newcommand{\tx}{Tx}

% Receiver
\newcommand{\rx}{Rx}

% Path Loss
\newcommand{\pl}{\Gamma}

% Transmit Power
\newcommand{\ptx}{P_{tx}}

% Receive Power
\newcommand{\prx}{P_{rx}}

% Ideal guessing bias
\newcommand{\biasid}{\rho}

% Non-ideal guessing bias
\newcommand{\biasnid}{l}

% Variance-based MTAC name
\newcommand{\varmtac}{Variance-Based MTAC}

% Distance
\newcommand{\dist}{d}

% False negative rate, #1: Alg name
\newcommand{\fnr}[1]{\mathit{FNR}_{#1}}

% Bit error rate
\newcommand{\ber}{\mathit{BER}}

% Frame error rate
\newcommand{\fer}{\mathit{FER}}

% Worst-case forger
\newcommand{\frgrnameworst}{\hat{\frgrname}}

% Receiver-side equivalent pulse train
\newcommand{\rxptr}{\mathbf{p}'}

% Null hypothesis
\newcommand{\nullhypo}{\mathcal{H}_0}

% Alternative hypothesis
\newcommand{\althypo}{\mathcal{H}_1}

%!TEX root =  mtac_main.tex
\section{Introduction}
When did the message arrive at the receiver? Can this estimate of the message arrival time be manipulated, and in particular by an attacker that controls the communication channel? In particular, can message advancement and delay attacks be prevented? This question is at the core of the problem that distance bounding protocols, secure positioning, and navigation systems are trying to solve: can we prevent the attacker from reducing or enlarging the distance that is measured between the devices? This problem is relevant in a number of application scenarios: contactless payments~\cite{markantonakis2012practical}, Passive Keyless Entry and Start Systems~\cite{francillon2011relay, ranganathan2012physical, mercedes_theft, hack_key}, GNSS (e.g., Galileo, GPS) security~\cite{humphreys2008assessing, papadimitratos2008gnss, tippenhauer2011requirements}. If we could prevent Time of Arrival (ToA) and therefore distance manipulation attacks, we could enable many proximity-based applications, from location-based access control to secure navigation~\cite{capkun2010integrity, rasmussen2007secnav}.

As a result, many distance bounding protocols have been proposed and analyzed~\cite{brands1993distance, kim2008swiss, brelurut2015survey}. Implementations of distance bounding protocols have emerged that combine such protocols with distance measurement techniques~\cite{Hancke_Kuhn_DB, rasmussen2010realization, tippenhauer2015uwb, ranganathan2015proximity}, in particular with UWB 802.15.4 radios~\cite{3db, deca, zebra}. 

The main idea behind these solutions is to prevent ToA manipulation by the randomization of message content. Namely, it was commonly believed that if the attacker cannot predict the bits of the messages, then he will not be able to advance their time of arrival at the receiver. In~\cite{clulow2006so} the authors argued this to be false -- since bits are encoded into symbols, attackers can advance their arrival time. Different physical-layer attacks followed also validating this in practice~\cite{flury2010effectiveness, poturalski2011distance, ranganathan2012physical}. This led to the conclusion that secure distance measurement systems can only be built with short symbols and using rapid bit exchange~\cite{clulow2006so}. Given the limits on the output power, such a result would mean that only short-range systems could be made secure. This was shown to be incorrect in~\cite{singhuwb}, which showed that longer symbols can be used if they are interleaved in transmission in a manner that is unpredictable to the attacker. This further demonstrated that secure, long-range distance measurement systems are possible. Recent works further show that, under certain conditions, distance enlargement can also be detected~\cite{singh2018UWB-ED}. All these works showed that consideration of the details of how bits are encoded into symbols (i.e., modulation) is crucial in the design of secure distance measurement systems.  

This discussion leads to the following questions:

\emph{Can we construct a generic message to symbol encoding that prevents any message advancement/reduction (and therefore distance delay/enlargement) for symbols of arbitrary lengths (and therefore arbitrary measurement ranges)?}

\emph{Can we derive the main principles for the design of such encodings?}

In this work, we show that answering these questions is indeed possible. To do so, we introduce Message Time Of Arrival Codes (MTACs), a new class of cryptographic primitive that allows receivers to verify if an adversary has manipulated the message arrival time. In a similar way that Message Authentication Codes protect message integrity, MTACs preserve the integrity of message arrival times. They are, therefore, fundamental to any protocol that relies on Time of Arrival information, such as clock synchronization~\cite{ganeriwal2008secure}, distance measurement~\cite{hu2003packet} and positioning protocols~\cite{capkun2005secure, capkun2006secure, sastry2003secure, singelee2005location}.

In the same sense that bits can be encrypted with a shared key, the shape of a signal can also be hidden by masking it with a random fast-changing sequence. However, to verify a signal shape, a receiver has to aggregate the signal over a considerable time interval in order to capture enough energy. This is especially so when sender and receiver are separated by longer distances.
If the attacker knows the temporal alignment of those aggregations with the signal, he can hide his guessing errors in the null space of the (linear) aggregation function.
Simple signal masking is, therefore, not sufficient for the protection against distance manipulation attacks. 
To address this problem, in addition to using cryptographically-secured modulation (i.e., signal generation), an MTAC also \emph{performs cryptographic checks of the consistency of the modulation} at the receiver.

We give a formal definition of MTACs and their security. We provide the main principles for the design of these codes. We review existing secure distance measurement schemes and draft standards and show how they fit within our MTAC definitions. We then introduce a new \emph{Variance-Based MTAC} that is inspired by our design principles. We show that adhering to these principles allows protection against physical-layer distance-reducing attacks over a wide, realistic performance region.
We systematically explore the trade-off between performance and security in ranging.

The rest of the paper is organized as follows. In Section~\ref{sec:background}, we introduce physical-layer attacks against distance measurement. Section~\ref{sec:mtac_def} then contains the formal security definitions. Section~\ref{sec:design_space} explores attack strategies. In Section~\ref{sec:existing}, we go over existing proposals. After that, we propose a Variance-Based MTAC in Section~\ref{sec:variance_based}, which we underline with simulations in Section~\ref{sec:analysis}. We conclude in Section~\ref{sec:conclusion}.

%!TEX root =  mtac_main.tex
\section{Background: Secure Distance Measurement}
\label{sec:background}
In this section, we introduce different RF techniques for distance measurement and highlight the challenges towards securing such systems against physical-layer attacks.

\subsection{Distance Measurement Techniques and Standards}
Establishing location or proximity both require estimating the physical distance between two or more wireless entities. Numerous wireless ranging and localization techniques have emerged in the last decade. Some of these observe physical properties of the signal such as RSSI~\cite{bahl2000radar} or phase~\cite{vasisht2016decimeter} that change as a function of propagation. However, both these properties can be controlled by an attacker that relays the signal and modifies them to fit another distance claim~\cite{truong2014comparing, olafsdottir2017security}. The only signal property that cannot be reliably controlled by an attacker is its time-of-flight (ToF). More precisely, an attacker cannot reduce ToF, as a signal cannot traverse space faster than the speed of light.

For ToF measurements, ultra-wideband impulse radio (UWB-IR) has emerged as a prominent technique for precise ranging. It allows high operating distances despite power constraints by transmitting multi-pulse symbols. UWB-IR ranging is in the process of being standardized in IEEE~802.15.4z and is becoming commercially available~\cite{3db, deca, zebra}.

\subsection{Distance-Bounding Protocols}
\begin{figure}
	\centering
	\includegraphics[width=0.95 \linewidth]{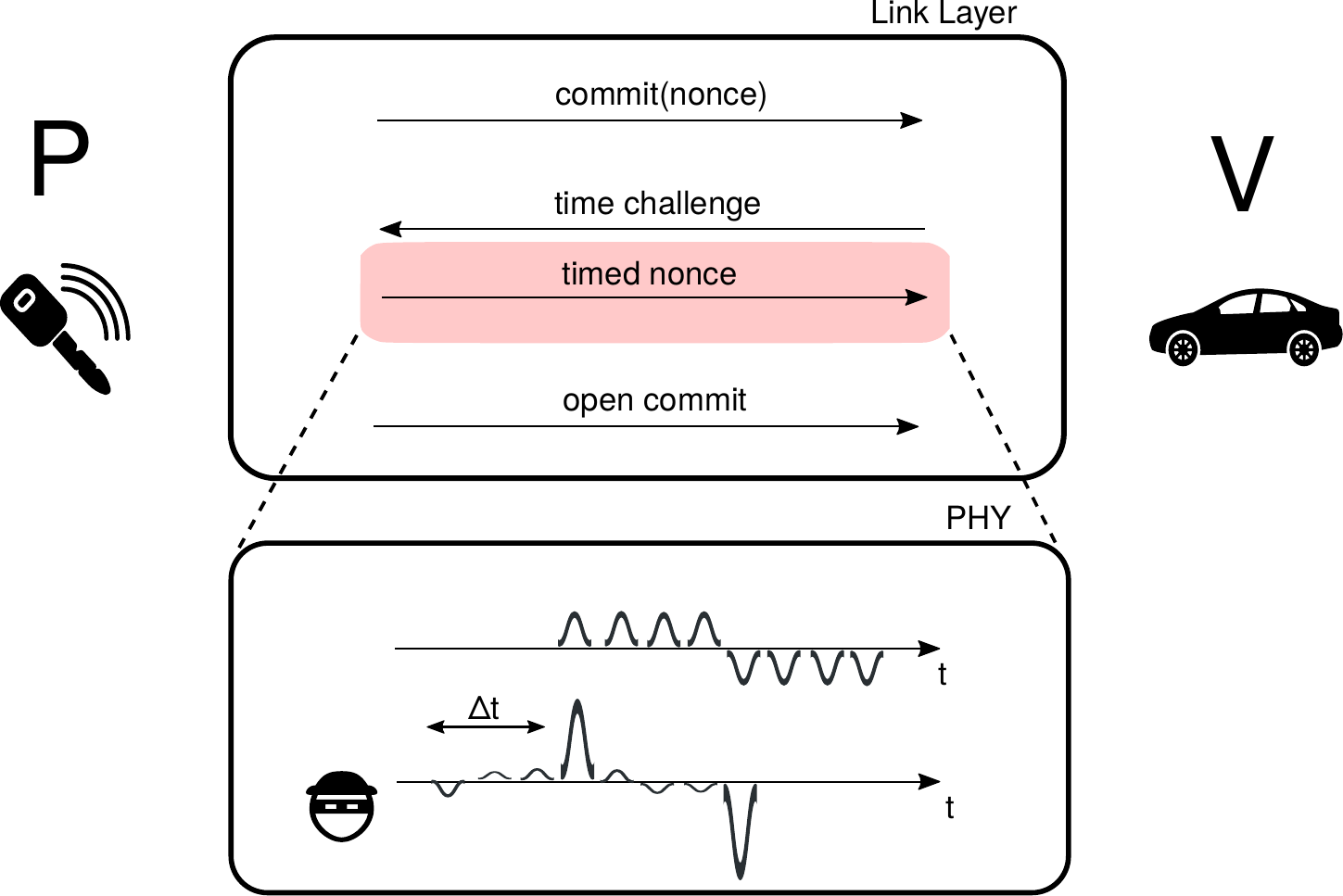}
	\caption{While distance-bounding protocols may be considered secure at the bit-level, systems can still be vulnerable at the physical layer. Distance-bounding protocols that rely on performant (i.e., long-distance), deterministic modulations are vulnerable to ED/LC attacks, as shown in the lower part of the figure. The underlying cause is time-redundant encoding for performance.}
	\label{fig:db_attack}
\end{figure}

Distance-bounding protocols that rely on ToF measurements, as provided by UWB-IR, are the cornerstone for secure proximity verification and positioning. As shown in Figure~\ref{fig:db_attack}, the basic idea behind such applications is as follows: a prover first commits to a cryptographic nonce; when triggered by receipt of a challenge message from the verifier, the prover sends the nonce, and then sends an opening of its commitment; verification is deemed successful if the commitment opens correctly to the nonce obtained at the verifier. ToF is bounded at the verifier by the time difference between sending the trigger signal and it starting to receive the nonce.

Existing vulnerabilities are related to the time-critical aspects of such a protocol, namely adjudging exactly when the nonce starts to be received at the verifier. This is relevant even when secret-dependent masked waveforms are used.
It is therefore essential that both the secret information content \emph{and} its time of arrival are carefully tested. Earlier instantiations of distance-bounding protocols relied on a rapid bit-exchange~\cite{tippenhauer2015uwb} to check both the content and timing of each bit of the nonce in consecutive rounds. As this requires each symbol to be short, this does not scale to longer distances. A distance commitment~\cite{tippenhauer2015uwb} can be used to decouple time acquisition from content verification. However, there are still doubts about the security level of the content verification, due to targeted attacks on the modulation~\cite{clulow2006so}.

\subsection{Physical-Layer Attacks}
Physical-layer attacks that target the underlying modulation cannot be addressed solely by distance-bounding protocols.
In the following, we do not consider attacks that can be averted by conservative signal acquisition (Cicada attack~\cite{poturalski2010cicada}) or involve denial of service (overshadowing, jamming). Instead, we address distance manipulation attacks that exploit redundancies in the modulation,  and that are not easily averted by security-aware configuration choices of existing receivers.

\subsubsection*{Early-Detect, Late-Commit (ED/LC) attack}
This attack reduces the distance measured by preemptively injecting a non-committal waveform that triggers an early signal detection at the receiver~\cite{clulow2006so, flury2010effectiveness}. The goal is to cause the receiver to register an earlier time of arrival, which, however, the attacker cannot back with knowledge about actual signal content. We illustrate this attack in the lower part of Figure~\ref{fig:db_attack}. The attacker gets away with this attack due to non-idealities of the legitimate receiver, requiring it to integrate signal power over time for each bit-wise decision, effectively limiting its resolution. To compensate for early deviations from the legitimate symbol (i.e., guessing errors), the attacker significantly amplifies his signal towards the end of each symbol. For maximum effect (distance reduction),  the attacker sends the committal, information-bearing part as late as possible after the start of the injected signal. Ideally, this is done to precisely coincide with the start time of the legitimate signal so that the attacker can ``copy'' it's content (with amplification). An ED/LC attack can be executed fully deterministically and can lead to a distance reduction up to the product of the symbol duration and the speed of light.

\subsubsection*{Guessing attacks}
If the polarity of individual pulses making up a modulated symbol does not only depend on the bit-value of the symbol, e.g., by being fully randomized as in~\cite{singhuwb}, the attacker can resort to a probabilistic ED/LC attack, i.e., a guessing attack. Here, the attacker tries to guess signal components in advance in order to reduce the measured distance. As in an ED/LC attack, the attacker exploits signal redundancies that are required for robust signal reception.
The basic idea is that the attacker can compensate early guessing errors by using more power towards the end of the symbol. For each symbol, the attacker can, for instance, double the power as long as his guesses are wrong and stop interfering as soon as a pulse is guessed correctly. This \emph{power-increase} attack is discussed in more detail in Section~\ref{sec:design_space}.

\subsection{Secure Distance Measurement Solutions}
There has been a proposal addressing the outlined threats by cryptographically hiding the bit-wise aggregations in a UWB-based On-Off Keying (OOK) modulation~\cite{singhuwb}. The authors provide concrete security levels for selected attacker models. A second approach is to correlate an incoming signal, a so-called Scrambled Timestamp Sequence (STS), with the expected signal shape and locking to the peak~\cite{4z_standard}. To the best of our knowledge, there exists no estimate of the concrete security offered by the latter method. In contrast to both proposals, our work establishes the security goal of any such approach on a fundamental level and outlines a solution permitting to a more general attacker and a broader performance region. We come back to the relation between our work and these existing schemes in Section~\ref{sec:existing}.

%!TEX root =  mtac_main.tex
\section{Message Time of Arrival Codes}
\label{sec:mtac_def}
In this section, we introduce Message Time of Arrival Codes (MTACs), physical-layer message codes that allow the receiver to verify the message time of arrival securely. Such codes preserve the legitimate signal time of arrival under an adversary that tries to ``shift'' the signal in time, i.e., aim to create the impression of a different arrival time. Fundamentally, the adversarial behavior can be either directed towards producing the code at a time earlier than its legitimate appearance (advancement) or to erase any evidence of a signal, thus opening the possibility for a late imitation (delay). In the following definitions, we divide the problem along those two threats. Although we provide definitions for both threats, we focus on their use in preventing message advancement (i.e., distance reduction) attacks. Preventing such attacks is necessary for the security of \emph{all} secure distance measurement systems. A system performing secure time of arrival measurement might in practice use separate codes to protect against advancement and delay attacks.

Prior to the definitions, we briefly go over boundary conditions on our proposal, stemming from requirements on precision and performance of wireless ranging signals.

\subsection{System and Attacker Model}
\label{sec:sys_att}

\begin{figure}
    \centering
    \begin{minipage}{0.22\textwidth}
        \centering
        \includegraphics[width=1\textwidth]{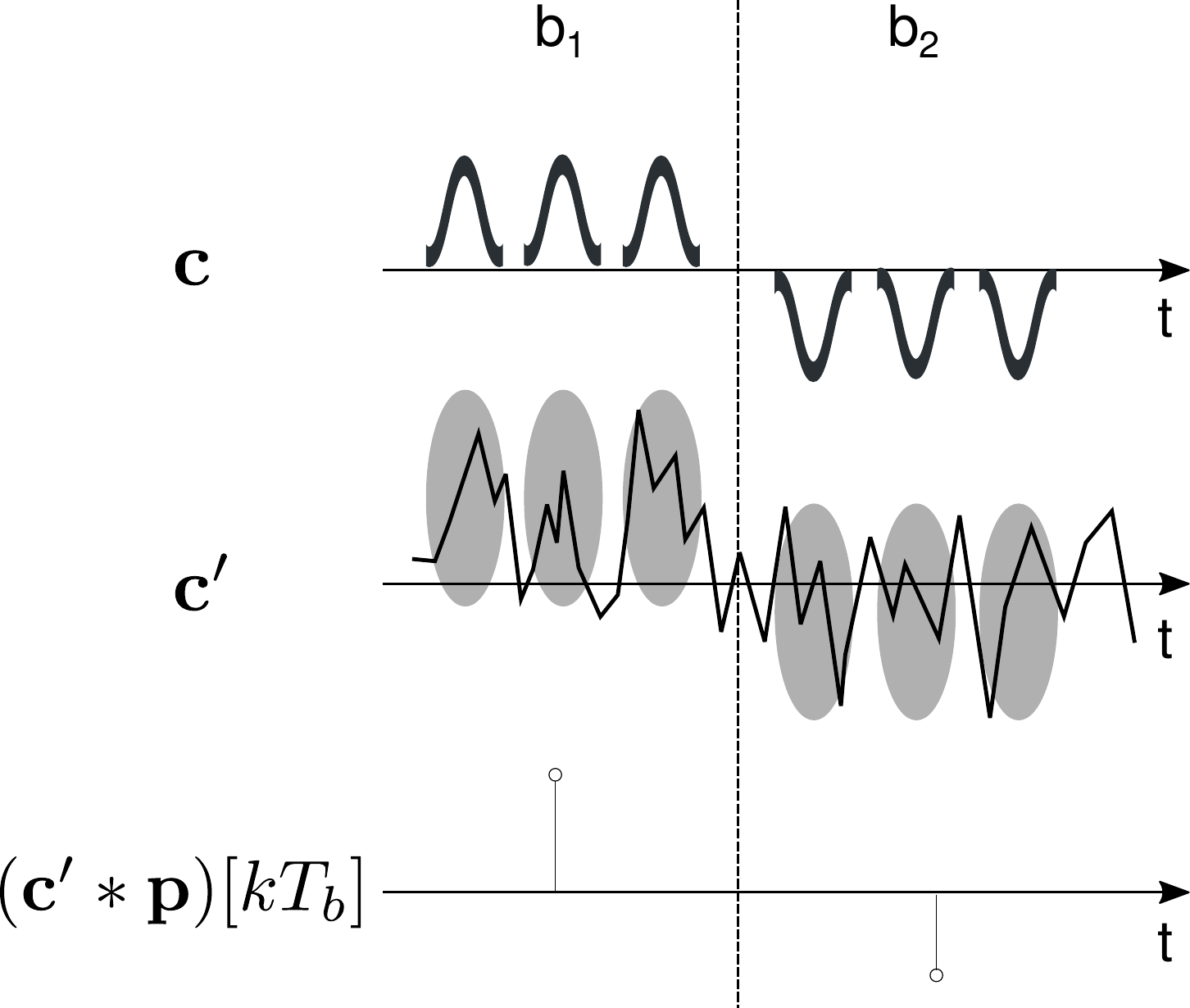}
    \end{minipage}\hfill
    \begin{minipage}{0.24\textwidth}
        \centering
        \includegraphics[width=1\textwidth]{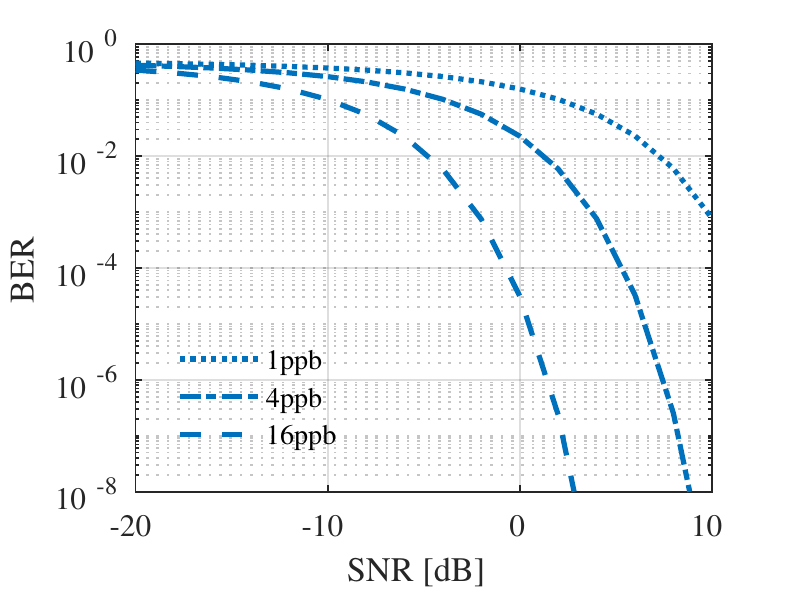}
    \end{minipage}
    \caption{Under noisy conditions, the receiver has to combine multiple short-term signal contributions (samples) to retrieve information. $(\rxsig * \mathbf{p})$ denotes linear aggregation, e.g., through a matched filter.}
    \label{fig:symbol_and_ber}
\end{figure}

We can model any ranging signal as consisting of short-time signal contributions (i.e., pulses) that carry the information used for precise ranging. As shown in Figure~\ref{fig:symbol_and_ber}, linear combinations of these pulses provide the statistics for the detection of information bits at the receiver. This model covers a broad range of modulation schemes.

\subsubsection*{Modulation}

In the following, we state some assumptions on the modulation.
Following Kerckhoffs' principle, we assume the attacker to be aware of all of these aspects of the modulation.
\begin{itemize}
\item The modulation consists of a series of elementary, short-time signal contributions called pulses. The effect of ED/LC attacks on such individual signal contributions is considered insignificant (say, less than 1m) in a sufficiently wideband system.
We refer to the amplitude level of such a pulse as a \emph{sample}.
\item For performance-related considerations, we assume the pulses to be sufficiently spaced such that there is no inter-pulse-interference in the given channel.
\item Each information bit is encoded in a symbol consisting of $\nppb$ pulses (and samples). The value of $\nppb$ is chosen in compliance with a target performance level $\perflvl$ within a performance region \mbox{$\perfreg=[0,\ber_{max}]\times[0,\dist_{max}]\times[0,\pl_{max}]$}, defined by intervals bounded by the maximally tolerated bit error rate $\ber_{max}$, the maximum communication distance $\dist_{max}$ as well as the maximum NLoS signal attenuation $\pl_{max}$.
\item Bits are grouped together to form frames, and each frame consists of $\frmlen$ pulses (and hence $\frmlen/\nppb$ bits).
\end{itemize}

\subsubsection*{Receiver Demodulation}
We assume the receiver demodulates by aggregating $\nppb$ samples using correlation with a polarity\footnote{Polarity refers to one of two possible phase values of the sample.} mask that fits the corresponding hypothesis for each possible value of the information bit. Then, a binary hypothesis test is applied to recover each bit. This is illustrated in Figure~\ref{fig:symbol_and_ber}. We assume an AWGN\footnote{Additive White Gaussian Noise.} channel model without inter-pulse interference. In general, the bit error rate (BER) at the receiver is therefore given by the tail bound on the Gaussian distribution, i.e.
\begin{equation}
\label{eq:ber}
\ber = Q\left(\sqrt{\frac{\nppb \prx}{\sigma_n^2}}\right),
\end{equation}
under Gaussian thermal noise with variance
\hbox{$\sigma_n^2 = bW \cdot N_0,$}
where $N_0$ is the noise power spectral density at room temperature, $bW$ is the system bandwidth and $\prx$ is the receiver-side signal power.
Figure~\ref{fig:symbol_and_ber} highlights the effect of larger $\nppb$ (longer symbols) on BER. This is to highlight the beneficial effect of temporal diversity on performance. Although Equation~\ref{eq:ber} refers to a BSPK modulation, this effect extends to other modulation techniques. We note that, within this model, for any channel and target BER, there exists an adequate symbol length and assume that the receiver chooses the symbol length accordingly. In this work, we do not assume any (error-correcting) coding.

\subsubsection*{Attacker Model} 
We assume that the attacker fully controls the communication channel and has no limitations on how fast she can process messages and react to them. She is, therefore, able to detect individual samples ideally. As a consequence, the attacker's information advantage increases as the channel for legitimate communication worsens, e.g., due to increased distance.
We consider two distinct attack models capturing distance reduction (message advancement) and distance enlargement (message delay). In the case of the distance reduction attacker, we pose no restriction on the attacker's abilities regarding the speed of computation,\footnote{Although MTACs can be constructed so as to be information-theoretically secure, most practical schemes will require that the attacker is computationally restricted.} location, or control of the communication channel (e.g., we give the attacker the ability to record and reactively inject messages on the channel with negligible delay). The only restriction that we pose is that the attacker cannot transmit information faster than the speed of light. The attacker's sampling rate needs to be sufficient to recover the signal. For an attack to be effective, we don't need to assume that the attacker has a higher bandwidth since we assume the attacker can precisely synchronize to the start of the signal. For the distance enlargement attacker, we assume that the attacker is constrained in terms of location, computation and control of the environment such that she is only able to block the reception of samples if she can anticipate their polarity.
However, this includes attackers that operate with multiple (smart) antennas or increase noise levels at the legitimate receiver

\subsubsection*{Visualising Attacks}
\begin{figure}
    \centering
    \begin{minipage}{0.22\textwidth}
        \centering
        \includegraphics[width=1\textwidth]{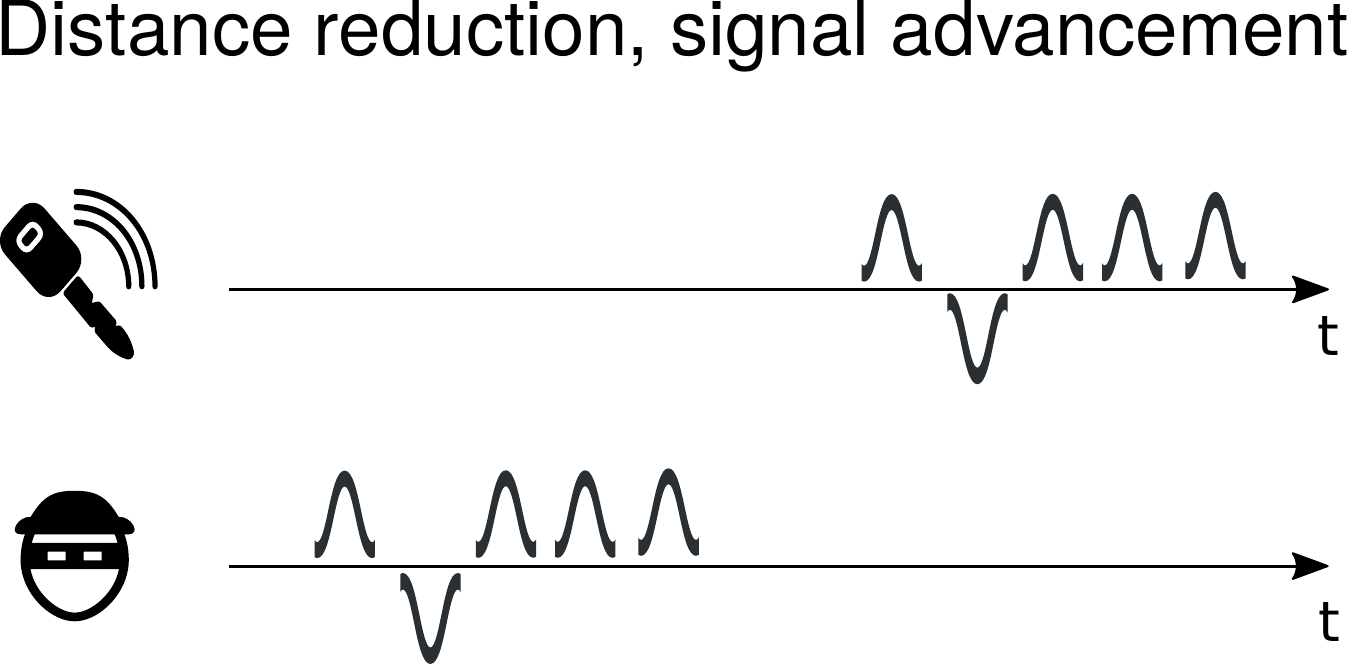}
    \end{minipage}\hfill
    \begin{minipage}{0.24\textwidth}
        \centering
        \includegraphics[width=1\textwidth]{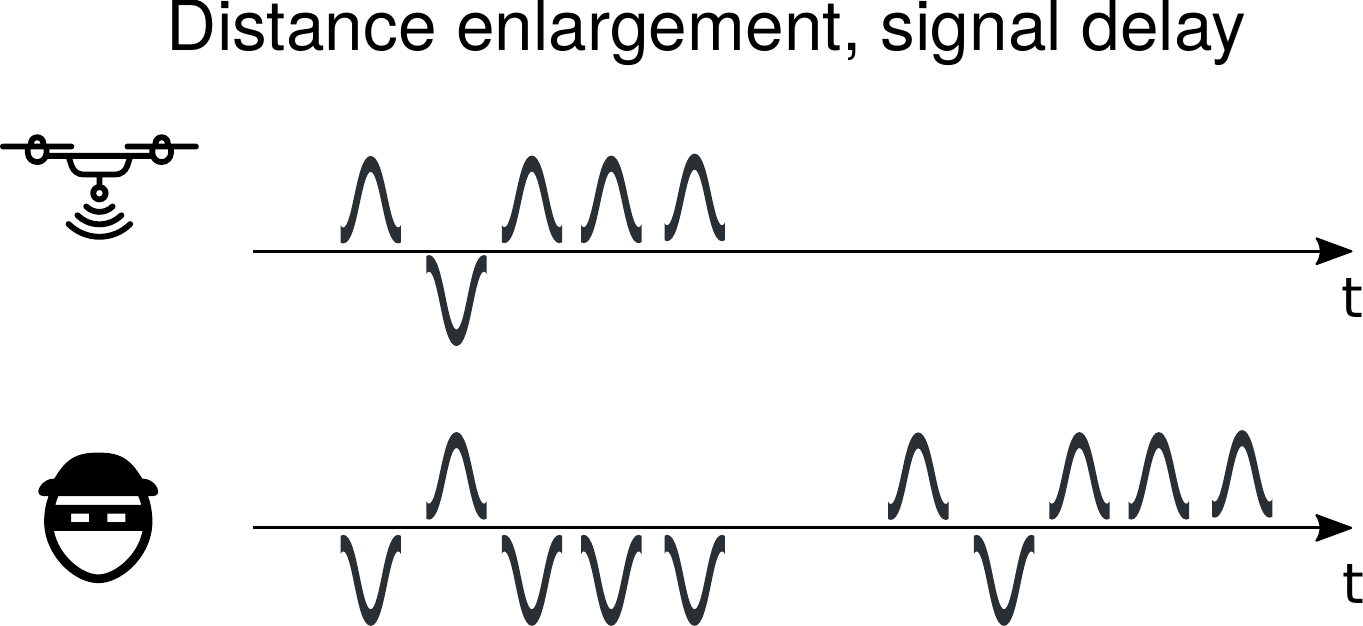}
    \end{minipage}
    \caption{Ideal instantiations of distance reduction (left) and distance enlargement (right) attacks. Both attacks are less likely to be detected the better an attacker can guess the legitimate signal shape. This holds both for releasing an early version (reduction) and for covertly annihilating the valid signal (enlargement).}
    \label{fig:attacks_ideal}
\end{figure}

In Figure~\ref{fig:attacks_ideal}, we illustrate ideal instantiations of distance modification attacks. Testing for a distance reduction attack at the receiver consists of a single hypothesis test: either the signal is real (i.e., \ only distorted by channel) or it is attacker-generated (i.e., \ it is distorted in a way that indicates that many pulses were guessed wrongly). An attacker is successful if he can produce the expected signal earlier. Verifying existence or absence of a distance enlargement attack, however, involves a multi-hypothesis test in time: the receiver has to check whether a given version of the signal is the first occurrence of its kind or if there exists an earlier, potentially degraded, version sufficiently similar to the legitimate signal. Consequently, in both attacks, an attacker's success chances are higher the better he can anticipate the legitimate signal shape.

\subsection{Definitions}
\label{sec:definitions}
A Message Time of Arrival Code (MTAC) is intended to allow detection of any kind of physical-layer distance-modifying attack with high likelihood.

\theoremstyle{definition}
\begin{definition}
A message time of arrival code (MTAC) is a tuple of probabilistic polynomial-time algorithms ($\genalgname$, $\mtacalgname$, $\vrfyalgname$), such that: 
\begin{enumerate}
    \item The key-generation algorithm $\genalgname$ takes as input the security parameter $\secprmlen$ and outputs a key $\key$ with $|\key| = \secprmlen$.
    \item The code-generation algorithm $\mtacalgname$ takes as input a key $\key$ and a message $\txmsg \in \{0,1\}^{\nbit}$ and outputs a real-valued vector $\txsig = (\txsigs{1}, \ldots, \txsigs{\frmlen})$. Since this algorithm may be randomized, we write this as $\txsig \leftarrow \mtacalgname_{\key}(\txmsg)$. 
    	\item The verification algorithm $\vrfyalgname$ takes as input a key $\key$, a real-valued vector $\rxsig$ of length $\frmlen$,
	and message $\rxmsg$. It outputs a bit $\vrfybit$. We assume that $\vrfyalgname$ is deterministic, and so write \mbox{$\vrfybit := \vrfyalg{\key}{\rxmsg}{\rxsig}$}.
\end{enumerate}
\end{definition}

In the above definition, we assume that $\txmsg$ may be transmitted separately from $\txsig$; however $\txsig$ can also `carry' $\txmsg$, which case we assume the existence of an efficient algorithm to extract $\txmsg$ from $\txsig$. In this situation, we can also assume that $\rxmsg$ can be extracted from $\rxsig$ and could choose to suppress it as an input to $\vrfyalgname$. The value of  $\vrfybit$ output by $\vrfyalgname$ is intended to convey that message time of arrival is correct ($\vrfybit = 1$) or that it cannot be securely verified ($\vrfybit = 0$).

An MTAC can be seen as a keyed signal verification scheme that guarantees the integrity of the message time-of-arrival. $\txsig=(\txsigs{1}, \ldots, \txsigs{\frmlen})$ is a vector of samples corresponding to the digital representation of the analog signal after A/D conversion.
We make no assumptions on the confidentiality or authenticity of $\txmsg$. We assume that these can be achieved through other means, e.g., using encryption or message authentication codes.

\begin{figure}
	\centering
	\includegraphics[width=0.766 \linewidth]{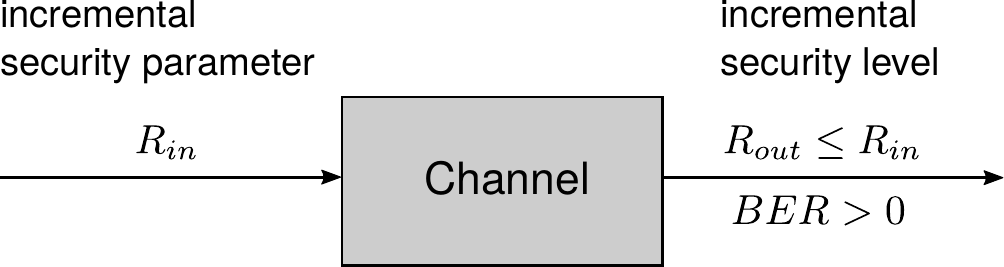}
	\caption{The wireless channel poses a fundamental indirection between the security parameter and the achievable security level. The detectable information rate at the receiver is smaller than the security parameter per second at the transmitter. The particular ratio $R_{out}/R_{in} = 1/n_{ppb}$ results from the modulation and reflects both a performance goal and the channel quality.}
	\label{fig:channel_loss}
\end{figure}

Before information can be verified, it has to be transmitted over a wireless channel and detected by the receiver. Strictly speaking, $\vrfyalgname$ involves not only verification but also time-selective \emph{detection} of physical-layer information. As highlighted in Figure~\ref{fig:channel_loss}, detection performance and the resulting security level are fundamentally connected. In general, received samples $\rxsig$ are affected by channel noise and, in consequence, not identical to $\txsig$. The detection rate $R_{out}$, which depends on channel and modulation, is the rate of verifiable information at the receiver. Due to temporal aggregation, it is, in general, smaller than the input data rate, i.e., $R_{out} \leq R_{in}$. Within our assumptions, the ratio $R_{in}/R_{out}$ is given by $\nppb$. Moreover, detection of this information over a channel is error-prone, which is reflected by a nonzero BER. Consequently, an MTAC will have a non-zero likelihood of false negatives, as well. This we address in a verification criterion that we call \emph{robustness}.
\theoremstyle{definition}
\label{def:robustness}
\begin{definition} An MTAC is robust if
\begin{enumerate}
\item In the absence of an attacker, for any channel, $\vrfyalgname$ applied on $\rxsig$ is falsely negative with probability at most $1-(1-\ber)^{\nbit}$, where BER is the error rate in detecting the bits carried by $\txsig$.
\end{enumerate}
\end{definition}
This means that the false negative rate should remain bounded by the frame error rate on the bit level. Note that we will impose robustness only on detection of distance advancement. As mentioned earlier, detection of delay attacks involves a multi-hypothesis test in time and is, therefore, inherently more prone to false positives.

Distance modification can mean either distance reduction or distance enlargement. The former requires the attacker to \emph{advance} the signal in time, the latter to \emph{delay} the signal in time. We define two different MTAC security models, one for each type of attack (a single model would be unwieldy and difficult to use).

\subsubsection*{\mtac{\advid}: Modelling Advancement Attacks}
In what follows, $\obsdel \ge 0$ denotes the \emph{observation delay} of the adversary, measured in samples, representing how long it takes for an attacker to observe and react to a given sample.\footnote{Although in our attacker model, we pose no restriction on the adversary's abilities to reactively record and inject samples, $\obsdel$ allows us to model weaker attackers whose reaction speed is bounded.} On the other hand, $\advgoal \ge 1$ denotes the number of samples by which the adversary tries to advance the signal, quantifying its attack goal.
Informally, we allow the adversary access to MTAC code values $\txsig$ for message inputs of its choice in a fully adaptive manner. 
Then we challenge it to produce an ``advanced'' signal $\rxsig$ for a message $\txmsg$ of its choice. We model the latter by requiring the adversary to produce component $\rxsigs{i+\advgoal}$ of its output before being given samples $(\txsigs{1},\ldots,\txsigs{i-\obsdel})$ of $\txsig=\mtacalgname_{\key}(\txmsg)$. The adversary wins if it eventually produces a vector  $\rxsig$ for which $\vrfyalgname_{\key}(\txmsg,\rxsig) = 1$. An MTAC scheme is (informally speaking) secure against advancement attacks if the probability that any efficient adversary wins is small. 

We formalise these ideas in terms of a message time-of-arrival forgery experiment $\mtacforgeKP{\advid}{\frgrname}{\algset}{\secprmlen}$. In this experiment:
\begin{enumerate}
    \item The experiment sets $\key  \leftarrow \genalgname(\secprmlen)$.
    \item The adversary $\frgrname$ is given oracle access to $\mtacalgname_{\key}()$; let $\qryset$ of size $q$ denote the set of queries made by $\frgrname$.
    \item $\frgrname$ outputs $\txmsg$, and the experiment sets $\txsig=\mtacalgname_{\key}(\txmsg)$.
    \item $\frgrname$ then sequentially outputs real values $\rxsigs{1}, \ldots, \rxsigs{\frmlen}$; however, after outputting $\rxsigs{i+\advgoal-1}$ (and before outputting $\rxsigs{i+\advgoal}$), $\frgrname$ is given the samples $(\txsigs{1},\ldots,\txsigs{i-\obsdel})$ of $\txsig$.
    \item Let $\rxsig$ denote $(\rxsigs{1}, \ldots, \rxsigs{\frmlen})$. Then the output of the experiment is defined to be $1$ (and $\frgrname$ is said to win) if and only if (1) $\vrfyalgname_{\key}(\txmsg, \rxsig) = 1$ and (2) $\txmsg \notin \qryset$. Otherwise, the output of the experiment is defined to be $0$.
\end{enumerate}

Note that for schemes in which a message $\rxmsg$ (possibly different from $\txmsg$) can be extracted from $\rxsig$, we can define a different win condition: (1) $\vrfyalgname_{\key}(\rxmsg, \rxsig) = 1$ and (2) $\rxmsg \notin \qryset$. Here, $\frgrname$ still outputs a message $\txmsg$ for which she receives a delayed version of $\txsig=\mtacalgname_{\key}(\txmsg)$, but she can win by ``forging'' a code vector $\rxsig$ for a different message $\rxmsg$ altogether.

\begin{definition}
Let $\algset$ = $\{\genalgname, \mtacalgname, \vrfyalgname\}$ be an MTAC-A, and let $\frgrname$ be an adversary with observation delay $\obsdel$ and advancement goal $\advgoal$ that makes at most $q$ queries to its MTAC oracle and that runs in time at most $t$ (across all steps of the  $\mtacforgeKP{\advid}{\frgrname}{\algset}{\secprmlen}$ experiment). The advantage of $\frgrname$ is then defined as: 
\[ \mtacadvKP{\mtac{\advid}}{\frgrname}{\algset}{\secprmlen} := \Pr[\mtacforgeKP{\advid}{\frgrname}{\algset}{\secprmlen} =1].\]
We associate with $\algset$ an insecurity function $\mtacadvmaxKP{\mtac{\advid}}{\algset}{\cdot}{\cdot}{\cdot}{\cdot}{\cdot}$, defined as:
\[
\mtacadvmaxKP{\mtac{\advid}}{\algset}{q}{t}{\obsdel}{\advgoal}{\secprmlen} := \max\limits_{\frgrname} \{  \mtacadvKP{\mtac{\advid}}{\frgrname}{\algset}{\secprmlen} \}\]
where the maximum is taken over all adversaries with observation delay $\obsdel$, advancement goal $\advgoal$, making at most $q$ queries to its MTAC oracle and running in time at most $t$. 
\end{definition}
It is not hard to see that, with all other parameters fixed, the insecurity function is maximised w.r.t. $\obsdel$ and $\advgoal$ when $\obsdel=0$ and $\advgoal=1$. This corresponds to the situation where the adversary has no observation delay and is given the next sample $\txsigs{i}$ from $\txsig$ immediately after outputting its own guess $\rxsigs{i}$. The latter corresponds to an adversary who tries to advance the signal by one pulse.

\begin{figure}
	\centering
	\includegraphics[width=1 \linewidth]{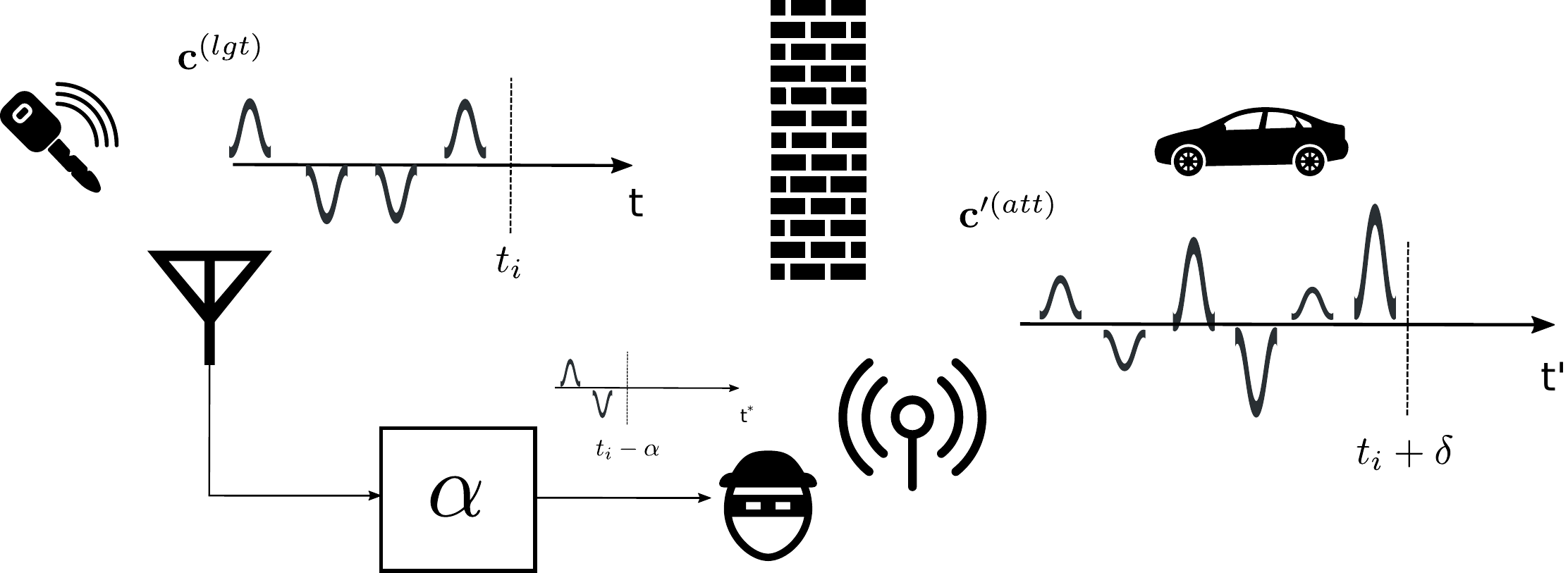}
	\caption{Distance reducing attack. The attacker sees the legitimate signal with an observation delay of $\obsdel$ samples and sends his guess $\advgoal$ samples ahead of the actual signal. If successful, the attacker can reduce the measured distance between key and car by $\advgoal$ samples.}
	\label{fig:attack}
\end{figure}

\subsubsection*{\mtac{\delid}: Modelling Delay Attacks}
In the following, we consider an adversary interested in removing all traces of the legitimate signal to perform a delay attack. Under the condition that all evidence of the legitimate signal is removed, the adversary can trivially achieve any delay goal $\advgoal$ without a risk of detection. As the value of $\advgoal$ does not help or limit the adversary, we are not using it in the model. However, by limiting the observation delay $\obsdel \ge 0$, we constrain the attacker in its ability to observe (and suppress) the samples that are transmitted by the legitimate transmitter. Generally, we assume that the attacker will not be able to detect the legitimate sample, transmit an opposite sample and thus suppress the legitimate sample.
%\textcolor{blue}{TODO: Limitations: Attacker in direct path, directionality (?).}
%$\obsdel$ in this scenario represents an attacker being located at some distance from the direct communication path.
Informally, we allow the adversary access to MTAC code values $\txsig$ for message inputs of its choice in a fully adaptive manner. Then, we challenge it to produce an ``advanced'' signal $\rxsig$ for the message $\txmsg$ of its choice.
%\textcolor{red}{This is advanced here as well. }    
We model the latter by requiring the adversary to produce component $\rxsigs{i}$ of its output before being given samples $(\txsigs{1},\ldots,\txsigs{i-\obsdel})$ of $\txsig=\mtacalgname_{\key}(\txmsg)$, i.e., the adversary needs to produce at least one sample in advance for $\obsdel = 0$. The adversary wins if it eventually produces a vector $\rxsig$ for which  $\vrfyalgname_{\key}(\txmsg,\spsig) = 0$ for $\spsig := \txsig + \rxsig$. $\vrfyalgname_{\key}(\txmsg,\spsig)$ outputs 0 if it does not find a trace of $\txsig$ in $\spsig$ and is unable to detect the existence of $\rxsig$.
%and $\rxsig$ is not a jamming signal \footnote{ I.e., the power is bounded by the given noise model}.
%, i.e, the values of $\rxsig$ is within the given noise distribution.

We formalise these ideas in terms of a message time-of-arrival forgery experiment $\mtacforgeKP{\delid}{\frgrname}{\algset}{\secprmlen}$ . In this experiment:

\begin{enumerate}	
    \item The experiment sets $\key  \leftarrow \genalgname(\secprmlen)$.
    \item The adversary $\frgrname$ is given oracle access to $\mtacalgname_{\key}()$; let $\qryset$ of size $q$ denote the set of queries made by $\frgrname$.
    \item $\frgrname$ outputs $\txmsg$, and the experiment sets $\txsig=\mtacalgname_{\key}(\txmsg)$.
    \item $\frgrname$ then sequentially outputs real values $\rxsigs{1}, \ldots, \rxsigs{\frmlen}$; however, after outputting $\rxsigs{i}$ (and before outputting $\rxsigs{i+1}$), $\frgrname$ is given the samples $(\txsigs{1},\ldots,\txsigs{i-\obsdel})$ of $\txsig$. Samples $\txsigs{i}$ and $\rxsigs{i}$ arrive at the receiver at same time, resulting in the superposition $\spsigs{i} = \txsigs{i} + \rxsigs{i}$.
    \item Let $c''$ denote $(\spsigs{1}, \ldots, \spsigs{\frmlen})$. Then the output of the experiment is defined to be $1$ (and $\frgrname$ is said to win) if and only if (1) $\vrfyalgname_{\key}(\txmsg, c'' ) = 0$ and (2) $\txmsg \notin \qryset$.
    Otherwise, the output of the experiment is defined to be $0$.
\end{enumerate}

\begin{definition}
	Let $\algset$ = $\{\genalgname, \mtacalgname, \vrfyalgname\}$ be an MTAC-D, and let $\frgrname$ be an adversary with observation delay $\obsdel$  that makes at most $q$ queries to its MTAC oracle and that runs in time at most $t$ (across all steps of the  $\mtacforgeKP{\delid}{\frgrname}{\algset}{\secprmlen}$ experiment). The advantage of $\frgrname$ is then defined as: 
	\[ \mtacadvKP{\mtac{\delid}}{\frgrname}{\algset}{\secprmlen} := \Pr[\mtacforgeKP{\delid}{\frgrname}{\algset}{\secprmlen} =1].\]
	We associate with $\algset$ an insecurity function $\mtacdelmaxKP{\mtac{\delid}}{\algset}{\cdot}{\cdot}{\cdot}{\cdot}$, defined as:
	\[
	\mtacdelmaxKP{\mtac{\delid}}{\algset}{q}{t}{\obsdel}{\secprmlen} := \max\limits_{\frgrname} \{  \mtacadvKP{\mtac{  \delid}}{\frgrname}{\algset}{\secprmlen} \}\]
	where the maximum is taken over all adversaries with observation delay $\obsdel$,  making at most $q$ queries to its MTAC oracle and running in time at most $t$. 
\end{definition}

With all parameters fixed, the insecurity function is maximized for  $\obsdel =0$. This corresponds to the situation when an attacker's observation delay is limited due to its position or hardware capabilities such that he cannot detect the legitimate sample and suppress them when they are already being transmitted. However, he can observe sample $\txsigs{i}$ from $\txsig$ immediately after outputting its own guess $\rxsigs{i}$.

Practical MTAC instantiations are likely to rely on a scheme to expand some finite sequence of ideal randomness into a longer one, e.g., using PRFs. We note that, in practice, this is the component vulnerable to higher values of $q$ and $t$. On the other hand, the security of the verification does not necessarily depend on $q$ and $t$, i.e., is not affected by those under the assumption of ideal randomness going into signal generation. This is equivalent to stating that verification is not necessarily randomized (beyond the randomness in the signal).
However, verification has to be reliable given some, within the computational model bounded, knowledge of the attacker about the PRF output used for signal generation.

%!TEX root =  mtac_main.tex
\section{MTAC Design Space}
\label{sec:design_space}

In this section, we shift to a statistical viewpoint on the design space of secure MTAC schemes and explain how this approach relates to the computational model presented earlier. A statistical analysis entails the advantage of summarizing the infinite number of possible attack strategies.
%In the following, we explore the space of strategies for distance-reducing attacks.
This is particularly beneficial because legitimate as well as adversarial signals can assume uncountably many realizations due to their real-valued nature and due to the uncertainty introduced by noise. Moreover, an attacker is free to choose any amplitude level for each sample of the transmitted signal.
The resulting complexity does not allow a straightforward evaluation of all possible strategies in a closed-form computational setting.
Also, the security of the verification procedure itself is best analyzed in information-theoretic terms, since verification itself does not have to be randomized, i.e., its security is not necessarily limited to a bounded adversary.
Therefore, we present a signal theoretic approach to evaluate different designs of MTACs and argue about the distinguishability of legitimate and attack signals in statistical terms.
Although such an approach does not support explicit bounds, we can encapsulate the infinite number of attack strategies and quantify their success in a holistic way. 
We compare different signals using both, distance on the bit level (Hamming distance) and distance on the sample level (\hbox{L2-distance}), which is motivated by the fact that attack success directly depends on the receiver's inability to distinguish an attacker's guessing error from noise.

Using our statistical model, we identify the symbol-wise mean\footnote{With mean we refer to the accumulated statistics per symbol after inner product with the expected polarity sequence.} and (residual) variance as the two main axes of optimization in any attack. We then derive meaningful over-approximations for these two properties that a successful attack signal needs to exhibit and 
define a strong attacker that will form the basis for the analysis in Section~\ref{sec:analysis}

\subsection{Distance-reducing attacker}
We ignore for a moment that the attacker has to provide a bit sequence that is accepted by the receiver and assume that the adversarial message passes bit-level verification.
In that case, detecting a distance-reducing attacker means distinguishing adversarial guessing errors from benign noise on the sample level.

To formulate such a test, we model noise and attacker error as stochastic processes $\nsprc$ and $\frgrsprc$. The noise process $\nsprc$ is i.i.d. Gaussian (AWGN channel), an assumption that holds as long as signal modulation places samples/pulses reasonably far apart to avoid inter-pulse interference.
The attacker process $\frgrsprc$, on the other hand, reflects the errors produced by the strategy to guess $\txsig$.
An attacker can freely choose the amplitude of its signal based on any strategy, however, $\frgrsprc$ is random w.r.t. the polarity of the adversarial samples since the attacker has to guess each sample of $\txsig$. We can capture this in the following hypothesis test:

\[\nullhypo :~ \ressig \sim \nsprc\]
\[\althypo :~ \ressig \sim \frgrsprc + \nsprc\]

For each time $j$ (corresponding to one sample), the noise process is distributed as $\mathbf{N}[j]\sim \nprc(0,\sigma_n)$, the attacker residual as $\mathbf{A}[j] \sim \mathcal{A}_j(\frgrname)$, for an attack strategy $\frgrname$.  The best strategy is the one for which the hypothesis test distinguishing $\mathbf{A}$ from $\mathbf{N}$ fails with the highest likelihood.

\emph{Together with the bit-level requirement that we have so far ignored, we can now formulate any attacker's universal goals as:}
\begin{enumerate}
	\item \textbf{Create the correct bits:} In order to achieve correct detection of each bit\label{itm:1}, the attacker needs to shift the signal mean $\mu_{\rxbit{i}}$ w.r.t. the polarity sequence of each symbol $i\in\{1,\ldots, \nbit\}$ beyond the sensitivity of the receiver.
	\item \textbf{Minimize the error energy:} The attacker aims to minimize the residual energy, i.e., the variance of his error distribution $\frgrprc_j$ at any time $j$\label{itm:2}.
	\item \textbf{Make the error as indistinguishable from noise as possible:} The attacker aims to hide in the noise the unavoidable\footnote{Since being related to the underlying hardness of guessing the pulses correctly.} guessing error, i.e., to bring the distribution $\frgrprc_j$ close to the legitimate noise distribution $\nprc (0,\sigma_n)$\label{itm:3}.
\end{enumerate}

Goal~\ref{itm:1} targets correctness on the bit level, whereas Goals~\ref{itm:2}~and~\ref{itm:3} are about indistinguishability of the guessed signal from the expected signal on the physical layer. As we will show, for Goal~\ref{itm:2}, there exists a clear relation to the hardness of guessing each signal sample of $\txsig$.

In the presented statistical model, achieving all three goals together represents a sufficient condition for attack success, irrespective of potential countermeasures (i.e., detection techniques).
There are different ways an attacker can go about these goals: an attacker can (1) select the subset of samples/pulses she wants to interfere with, (2) choose arbitrary amplitude levels for each targeted pulse, and (3) decide how many samples need to be observed before interfering. A meaningful attack strategy will be concerned with how to make these choices in order to satisfy all three goals jointly.

\noindent We now describe two general concepts that guide any attack strategy and lead to the definition of a strong attacker by over-approximating signal mean and residual energy.

\begin{figure}
	\centering
	\includegraphics[width=0.5 \linewidth]{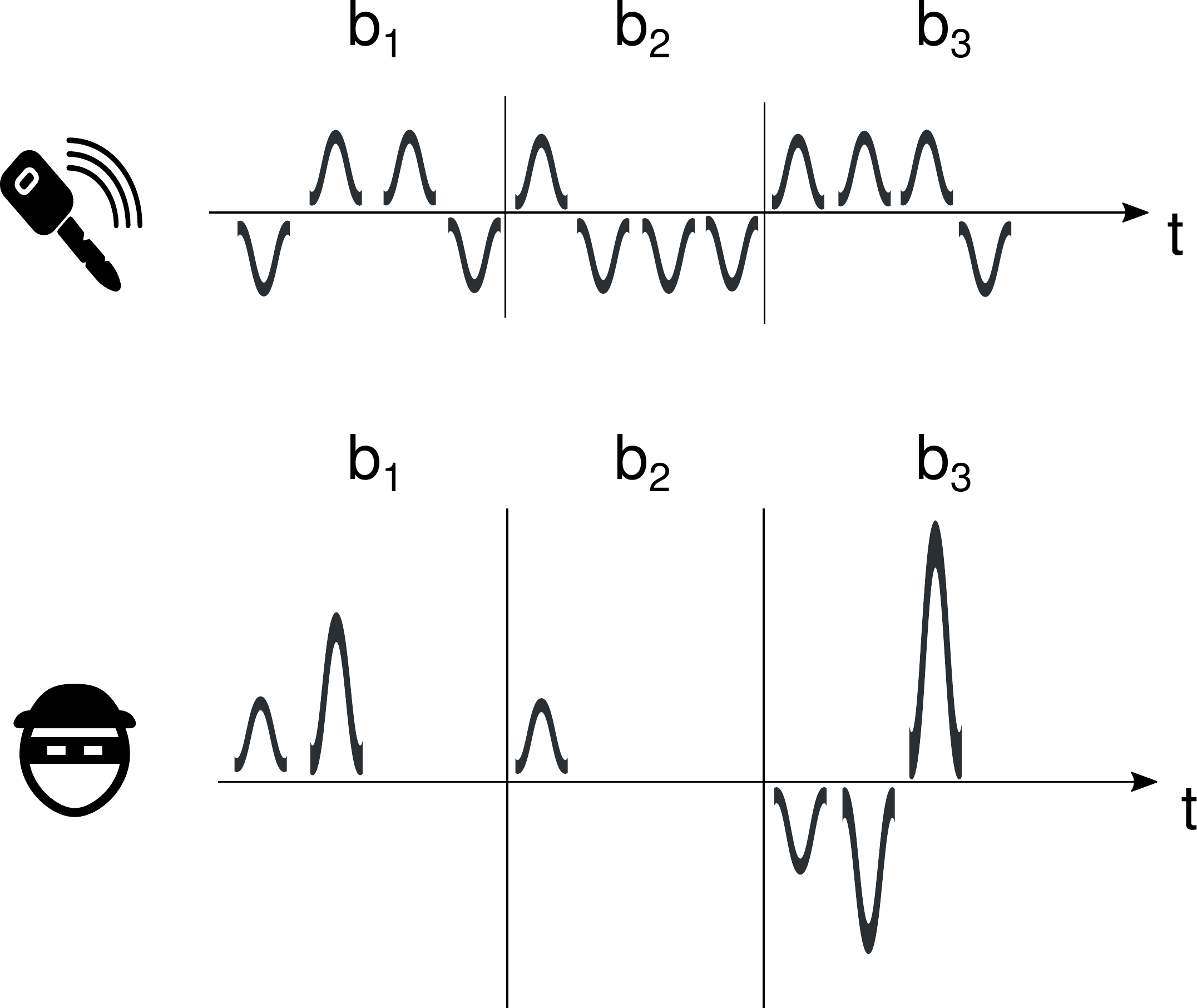}
	\caption{Even under a fully randomized pulse sequence holds: If the receiver (i.e., verifier) combines the pulses to symbols in a predictable manner, the attacker has high chances of getting a sufficiently high symbol-wise mean, by increasing the power in reaction to wrong polarity guesses.}
	\label{fig:power_increase}
\end{figure}

\subsubsection*{Steering the mean: Power-increase strategy}
Even if the signal is fully randomized at the pulse level,  an attacker can guess symbols by employing a \emph{power-increase strategy} as shown in Figure~\ref{fig:power_increase}. Fundamentally, pulse level randomization under sample-level feedback does not keep an attacker from steering his signal to an arbitrarily high mean under inner product with the hidden polarity sequence.
An attacker starts by sending a pulse containing the entire symbol power. He will keep on doubling the power per pulse until he guesses a pulse of the symbol correctly. This attack succeeds with probability $1-0.5^{\nppb}$ per symbol.
The core takeaway from this attack is that a sample-level guessing error of the attacker does not necessarily translate to a bit-level error, due to the dimensionality reduction applied at the receiver. As long as the attacker can hide the error in the null space of this linear transformation, there is no incentive against the attacker using progressively higher energy levels to 'force' the bits. This means, Goal~\ref{itm:1}, in isolation, is easy to achieve for an attacker. However, achieving the goal with high likelihood, i.e., more attempts, is associated with higher power levels, which puts Goals~\ref{itm:2}~and~\ref{itm:3} in increasing jeopardy.

\subsubsection*{Minimizing guessing error by learning pulse polarities}
Goals~\ref{itm:2}~and~\ref{itm:3} are directly related to the pulse-guessing performance of the attacker.
Depending on how the information bits are modulated, the attacker can potentially use bit-level information to infer the signal or rely on knowledge of past pulses to anticipate the pulse polarities ahead. This would reduce the guessing error and make it harder to detect the attack. Our attacker, as introduced in Section~\ref{sec:definitions} has full knowledge about the transmitted bits.
In general, any unmasked signal redundancy in time can potentially help the attacker. An example of this is repetition coding or bit-level error-correction coding (ECC) as used in the coherent mode of IEEE~802.15.4z~HRP~\cite{4z_standard}. Also, nonidealities in the underlying PRF can help an attacker.

\subsubsection*{A strong attacker}
\begin{figure}
	\centering
	\includegraphics[width=0.6 \linewidth]{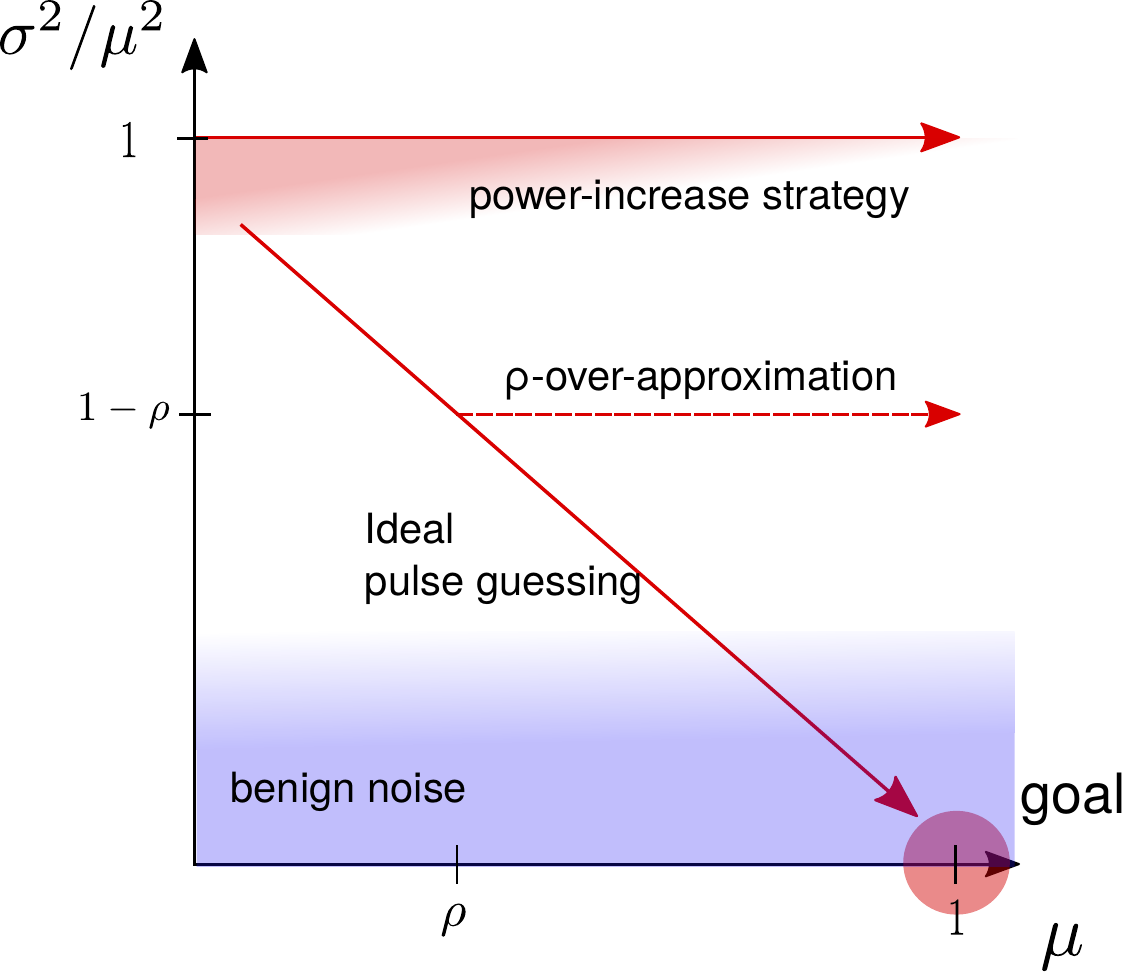}
	\caption{Attacker's strategy space. An attacker needs to exceed a certain symbol-wise mean to produce the correct bits a the receiver. This he can achieve with high likelihood using a power-increase strategy. However, there does not exist any reliable strategy for decreasing the normalized error variance. An attacker can only do so by maintaining an edge in guessing pulse polarities. This we model by over-approximating the attacker, e.g., by giving him a pulse-guessing bias $\rho$.}
	\label{fig:strategy_space}
\end{figure}

We abstract away from all possible strategies and only describe the attack signal statistically subject to an over-approximation of its properties that are linked to the attacker's success: signal mean\footnote{i.e., the inner product with the expected polarity sequence. Correct guesses contribute to it, wrong guesses diminish it.} and residual energy (i.e., residual variance).

As will be motivated, residual variance emerges as observable under a maximum entropy assumption on the attacker's strategy.
A result from information theory states that the Kullback-Leibler divergence (i.e., relative entropy) determines the exponent of the error in distinguishing two statistical distributions~\cite{cover2012elements}. Consequently, an attacker that brings its residual closest to the legitimate signal is the strongest. Therefore, we can define the strongest attacker $\frgrnameworst$ as the one that is closest in the KL-sense over all times:

\begin{align*}
	\frgrnameworst & :=  \argmax_{\frgrname} \adv (\frgrname) \\
	& = \argmin_{\frgrname} \sum_{j=1}^{\frmlen} D_{KL}(\frgrprc_j (\frgrname) + \nprc \ \Vert \ \nprc) \\
	& = \argmin_{\frgrname} \min_{j} \frmlen\  D_{KL}(\frgrprc_j(\frgrname) + \nprc \ \Vert \ \nprc) \\
	& = \argmin_{\frgrname} D_{KL}(\frgrprc (\frgrname) + \nprc \ \Vert \ \nprc)
\end{align*}
The strategy that produces the smallest statistical distance at any $j$ can be converted into the best strategy over the entire signal, by applying the same technique at any other time, since the noise is i.i.d. Therefore, we argue that the attacker that is locally optimal at any time is also optimal over the entire process. The strongest attacker is, therefore, the one that can produce a residual distribution $\frgrprc + \nprc (0, \sigma_n)$ that has smallest relative entropy compared to the legitimate noise distribution $\nprc (0,\sigma_n)$.
Under the condition that the attacker's error has nonzero energy, the process $\mathcal{A}$ that minimizes relative entropy to the AWGN only is also a Gaussian.

Therefore, as an over-approximation, we can model the attacker residual signal process as normally (i.e., maximum entropy) distributed stochastic process with zero mean and a variance given by the pulse-level guessing performance, which we over-approximate. This is equivalent to assuming maximum ignorance about the attacker's process beyond the existence of some residual energy. Under these conditions, e.g. from~\cite{lapidoth2017foundation}, we know that the signal energy is a sufficient statistic for distinguishing two i.i.d. $\nprc (0,\sigma_1)$, $\nprc (0,\sigma_2)$-distributed processes.

\begin{observation}
	The signal residual variance constitutes a sufficient statistic for detection of a guessing attack with a maximum-entropy residual under AWGN noise.
\end{observation}
Basing the classification on the residual energy is optimal if we can extract the attacker's error perfectly and within the assumptions, we can universally impose on the attacker's error process (i.e., being close to satisfying the three goals). A practical attacker will likely deviate from these assumptions, but in ways that \emph{add} distinctive properties (i.e., non-zero higher moments) to the residual distribution.
Conversely, an attacker that gets mean and variance right will win.
\begin{observation}
	The attacker getting the mean per bit right \emph{and} minimizing signal residual variance together constitute a sufficient condition for attack success.
\end{observation}

\begin{figure}
	\centering
	\includegraphics[width=1 \linewidth]{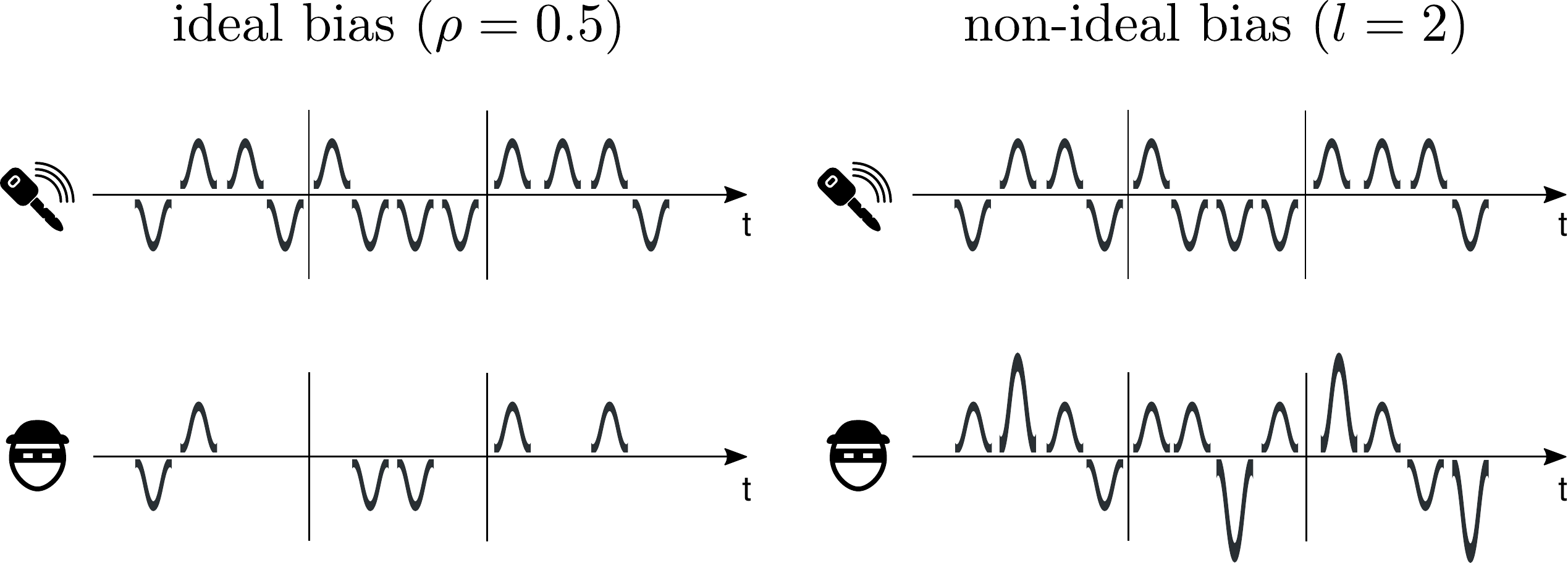}
	\caption{We model two different over-approximations for the attacker's error variance level: An ideal bias, where an attacker knows a fraction $\biasid$ of the pulse polarities and a non-ideal bias, where we give the attacker a bound $\biasnid$ on the number of power levels for a successful power-increase attack.}
	\label{fig:over_approx}
\end{figure}

We have seen that, without countermeasures, a power-increase strategy leads to a guessing bias in the receiver-side security parameter (i.e., the bits). As an over-approximation for the course of a power-increase strategy, we can tilt the guessing performance in the attacker's favor \emph{on the pulse level}.
For instance, we can assume that the attacker never makes a wrong guess twice in a row. This means, after at most two interferences (i.e., pulses), the attacker is guaranteed to have made a positive net contribution to the receive statistics. We refer to this attacker as having a \emph{non-ideal bias}
of $\biasnid = 2$ and illustrate it in Figure~\ref{fig:over_approx}. There, we contrast it to an \emph{ideally-biased} attacker, which knows a given fraction $\biasid$ of pulses.

In Figure~\ref{fig:strategy_space}, we highlight the two-dimensional nature of the attack strategy. It is easy for an attacker to steer the mean by varying his energy levels, i.e., to move along the x-axis. However, he cannot control the error variance at the same time. So, any practical attacker strategy will be concerned with trading off those two goals. Providing the attacker an ideal bias results in a diagonal towards the desired spot of high mean and low variance. In addition, as part of any over-approximation, we assume the attacker to be successful regarding the mean (e.g., through a power-increase strategy). This means the attacker can move arbitrarily on the x-axis. In the following, we motivate a specific over-approximation for the error variance, i.e., the attacker's position on the y-axis.

\begin{observation}
For an attacker, reducing the signal error variance, while increasing its mean, is 'pulse-guessing-hard'. This means, without a systematic guessing bias, the (normalized) error variance is bound to increase in a guessing attack.
\end{observation}

\begin{figure}
    \centering
    \begin{minipage}{0.24\textwidth}
        \centering
        \includegraphics[width=1\textwidth]{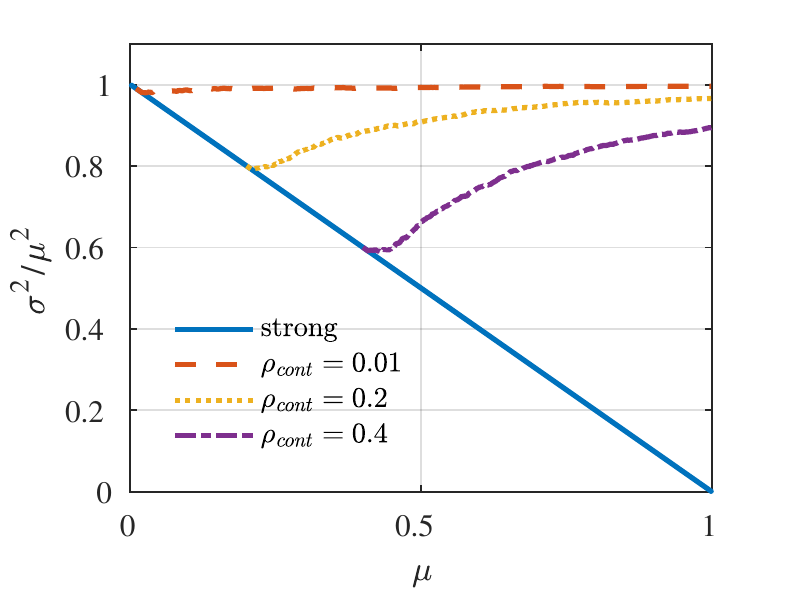}
    \end{minipage}\hfill
    \begin{minipage}{0.24\textwidth}
        \centering
        \includegraphics[width=1\textwidth]{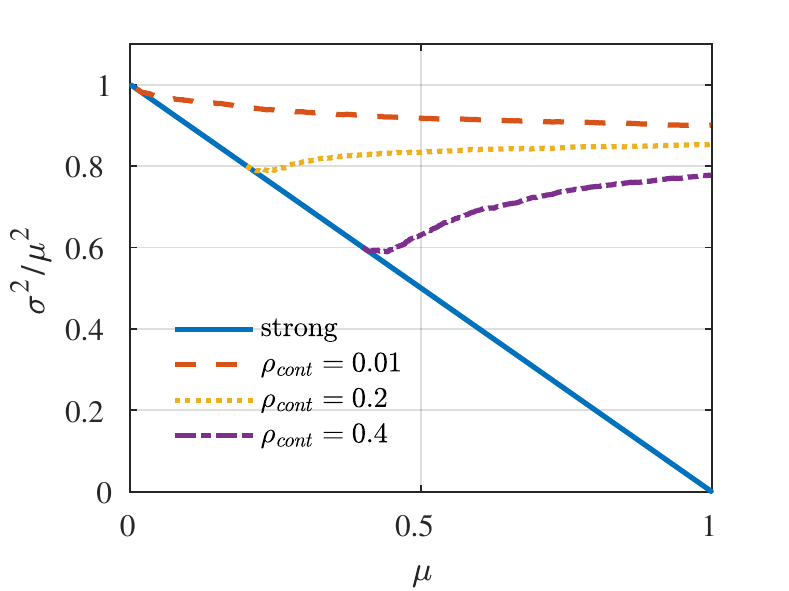}
    \end{minipage}
 		\caption{Normalized error variance vs. mean under over-approximation (blue) and continued interference (dashed lines). The goal of the attacker is to get mean to 1 while minimizing variance. Given an ideal bias above a certain threshold ($\biasid_{cont} \approx 0.2$), an attacker has nothing to gain from continued interference. The dashed lines show the 0.1th percentile of the variance for unbiased (left) and non-ideally ($\biasnid = 2$) biased (right) guessing continuation.}
    \label{fig:cont_if}
\end{figure}

In Figure~\ref{fig:cont_if}, we display simulation results underlining this. The results show the normalized residual energy of an ideally-biased attacker (blue line) as a function of the number of interferences, as well as the effect of continued interference without bias (left) and with a non-ideal bias (right). Without bias, the normalized variance is (mostly) monotonically increasing, converging to its maximum value of 1. With a non-ideal bias, the gain that can be maintained is limited. Even with such a consistent bias, only at low values for $\biasid$ is there any incentive to continue interfering. Especially, for $\biasid > 0.2$, there is no incentive to continue, even with a consistent but non-ideal bias.

\begin{observation}
Once the attacker has succeeded in shifting the mean for all symbols, there is (almost) no utility in continued interference, unless the attacker has a lasting pulse-guessing bias. But even then we can find an ideal bias $\biasid_{cont}$, such that there is no utility.
\end{observation}

We see in Figure~\ref{fig:cont_if} that persistent interference with a non-ideal bias alone (i.e., no ideal bias, red curve) results in a normalized variance of more than 0.8. We can estimate the strength of this over-approximation as $0.75^{\frmlen / 2}$.
This results from the fact that for every two pulses guessed by an attacker, we omit the possibility of two wrong guesses, an event with probability $0.25$. By comparing this value to the bit-equivalent MTAC security level of $2^{-\nbit}$, we can see that an over-approximation with $\biasid = 0.2$ is actually stronger than the bit-equivalent MTAC target security level for modulations with $\nppb > 2\frac{\log(0.5)}{\log(0.75)}\approx 4.82$, i.e., at least five pulses per bit. A decrease of the relative variance to 0.8 or, equivalently, an ideal bias of $\biasid = 0.2$ are, therefore, very strong over-approximations, i.e., on the order of the (receiver-side) security parameter, that become even stronger (less likely) for modulations over longer communication distances.

%!TEX root =  mtac_main.tex
\section{Existing  MTACs}
\label{sec:existing}

Based on our insights on the attack, we need an MTAC to verify the physical-layer integrity of a signal by measuring the (normalized) signal residual variance. To the best of our knowledge, there are three existing classes of MTACs that, as we argue, aim for this implicitly. Each class is parametrized by a performance parameter that allows to trade off performance and security. Note that, in the following, the robustness definition does not directly apply to the first two classes since those do not entail reliable information transmission.

\subsection{Sequences of single-pulse bits}
To allow for longer range while using short symbols, one could encode each bit as a single pulse and tolerate up to a certain pre-configured rate of bit errors $\thrs{\ber}$ in the verification step, as is currently proposed in 802.15.4z~LRP~\cite{4z_standard}. This results in a secure MTAC under the condition that the message $\txmsg$ is pre-shared between transmitter and receiver. Since relying on a single pulse makes bit transmission unreliable,
this is a purely physical-layer construct and does not allow for integrity-protected data transmission. In particular, a MAC will fail if there is a nonzero number of expected bit errors per message.
The resulting security and performance level both depend on the $\ber$ tolerance level.
For $\txsig \leftarrow \txmsg ~\oplus~ \xorseq$, where $\xorseq$ is an ideal random sequence, the attacker's advantage can be estimated using Sanov's theorem, provided in~\cite{cover2012elements}, as
\[P(\overline{X}_{\frmlen} \ge (1-\thrs{\ber})) =  2^{-\frmlen D_{KL}(P||S)}.\]
Here, $P$ and $S$ capture the empirical and theoretical binomial distributions, with $P = (1 - \thrs{\ber}, \thrs{\ber} ) $ and $S= (0.5,0.5)$, respectively. The random variable $\overline{X}_{\frmlen}$ denotes the number of bits guessed correctly by the attacker.
For $\thrs{\ber}<1$, we can achieve any concrete security goal by setting $\frmlen$ appropriately. For example, under the assumption of ideal randomness and an unbounded adversary, for $\obsdel=0$ and $\advgoal=1$, we achieve 32 bits of security by transmitting 116 bits while tolerating up to $\thrs{\ber}=20\%$ bit errors. These results directly translate to a computational setting with reduced shared randomness by replacing the ideal randomness with the output of a PRF.

\subsection{Correlation sequences}
A standard way of signal time acquisition is correlating an incoming signal with the expected signal shape and locking to the peak. This is suggested in current 802.15.4z~HRP standardization efforts by means of a so-called Scrambled Timestamp Sequence (STS)~\cite{4z_standard}. One could argue that secure ToA verification could be achieved by checking for this peak. However, we know that tests for time acquisition and content verification should not be naively coupled~\cite{poturalski2010cicada}. If only used for content verification, the security of such an approach depends on the detailed test the receiver applies, i.e., the degree to which the relative quality of the peak (peak-to-average ratio) is considered. The power level at the peak itself can be easily steered to the desired value by an attacker relying on a power-increase strategy. Therefore, tests based on a) the existence of a correlation peak or b) the peak power level are not secure. Further exploration of this technique is deferred to future work.

\subsection{Hidden encodings: UWB-PR}
Secure ToA-verification based on the bit-level content can be achieved by hiding the mapping from pulses to bits as in~\cite{singhuwb}. This can be thought of a scheme that reorders the pulses belonging to each bit within the frame. This is currently the only scheme that is implemented and secure. It forces the adversary to transmit a small number of pulses, and, in case their polarity is correct, hope that they align with the bits. The authors analyze the performance and security of such an attack model. The adversary is given the same capabilities as assumed in Section \ref{sec:sys_att}. They have analyzed attack strategies for $\obsdel=0$  and $\advgoal=1$, and have assumed that the communicating parties share large amounts of ideal randomness. However, a formal proof is needed to determine if the attack strategy is optimal. The results show that 16 pulses per bit are required under LoS conditions to prevent bit errors over a distance of 92m. For 32 bits of security at least 100 bits have to be sent in a message.
As long as the attacker does not guess all but $\nppb$ pulses correctly, he has an advantage less than one, due to his uncertainty about the reordering. Consequently, under this assumption, we can add more bits to the frame to achieve any concrete security level.

\section{\varmtac{}}
\label{sec:variance_based}

In the following, we propose the \varmtac{} for direct variance estimation, consisting of rules for signal creation and a receiver-side verification procedure.
%This allows us to move from asymptotic to concrete security guarantees.
We then embed this technique into a generic verification algorithm and address side requirements for its practical instantiation.% and discuss the differences of our proposal to existing concepts.

\subsection{Tx-side signal generation ($\genalgname$, $\mtacalgname$)}

We assume each sample to follow a binary encoding, achieved either through on-off keying (OOK), frequency-shift keying (FSK) or phase-shift keying (PSK), but not pulse-position modulation (PPM). The reason is that, in PPM, the fundamental signal contribution representing each sample becomes vulnerable to ED/LC.
Within our assumptions about the modulation,
we can represent the transmit signal as a binary pulse sequence of length $\frmlen = \nppb \cdot \nbit$. In particular, we assume that pulses are separated by more than the channel delay spread, i.e., there is no inter-pulse interference. Without this assumption, signal degradation under benign conditions might be hard to distinguish from attacks.
The bits are first encoded in a frame $\symbfrm = (\symb{\txbit{1}} \Vert \ldots \Vert \symb{\txbit{\nbit}}),$ consisting of symbols that each represent message bit under repetition coding, either as $\symb{1} = \{1\}^{\nppb}$ or $\symb{0} = \{-1\}^{\nppb}$.
Preventing an attacker from inferring pulse polarities from either the content of the message $\txmsg$ or past samples is achieved by relying on full pulse-level randomization, i.e., by applying a secret sequence $\mathbf{x}$ on the pulses, as in

\[\txsig = \symbfrm \oplus \xorseq.\]

We can either idealize $\xorseq$ being perfectly random, as in
\[\xorseq \leftarrow \{-1,1\}^{\frmlen},\]
and shared between transmitter and receiver, or being generated using a pseudorandom function that operates on a previously shared secret.

\begin{figure}
	\centering
	\includegraphics[width=1 \linewidth]{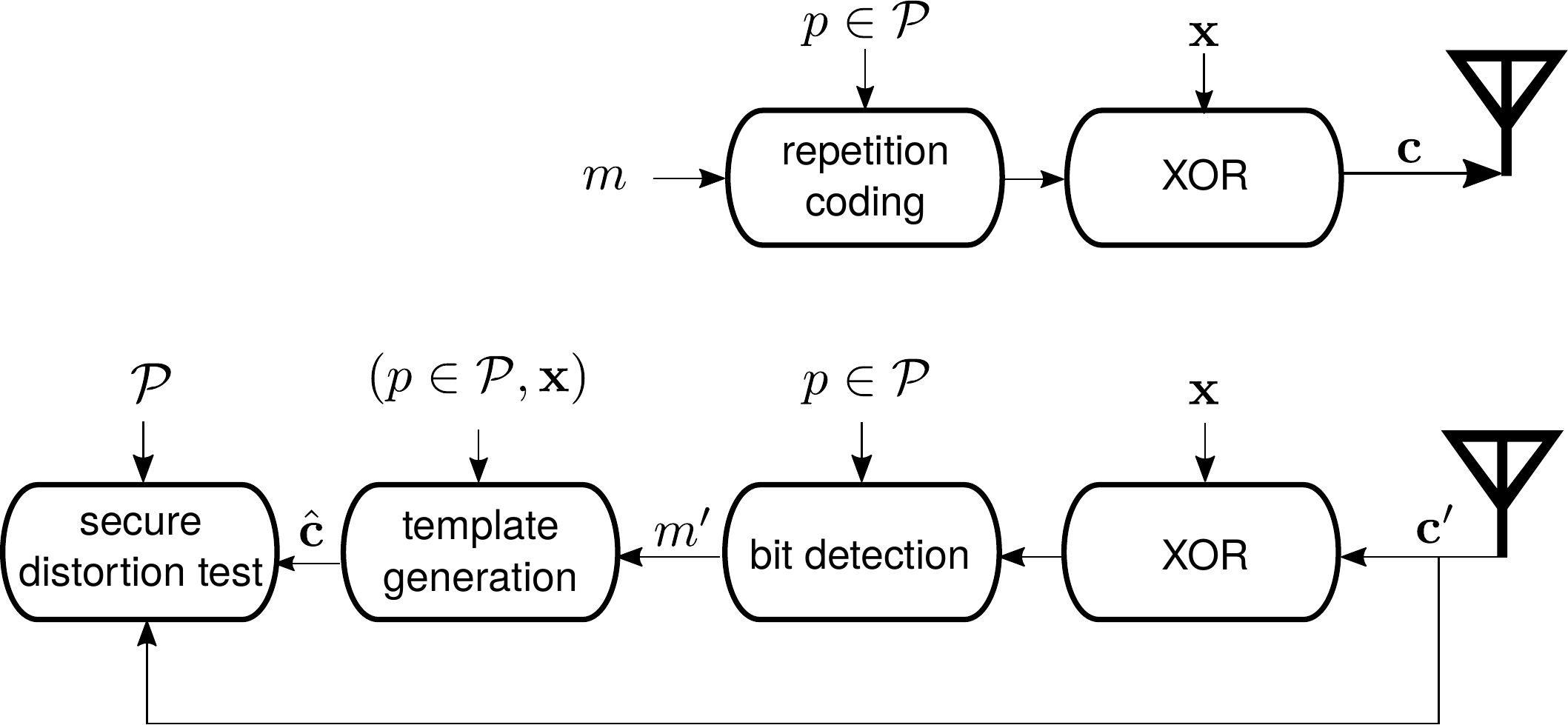}
	\caption{Tx/Rx structure of a \varmtac. A keyed XOR and a secure distortion test are the central security components. For simplicity, we omit the modulation of each value in $\txsig$ onto a UWB pulse in the picture. Bit encoding and decoding are parametrized by a performance level $\perflvl$, whereas the secure distortion test applies to an entire performance region $\perfreg$.}
	\label{fig:pipeline}
\end{figure}

\subsection{Rx-side operations ($\vrfyalgname$)}
A message time of arrival code has to combine bit detection and verification with an additional signal verification for ensuring the correct signal time of arrival. The bit-level tests are a sequence of binary hypothesis tests. The additional check is a single binary test applied to the entire signal, parametrized by the bits received. We illustrate the whole pipeline in Figure~\ref{fig:pipeline}.

\subsubsection*{Bit detection}
Each bit is carried by $\nppb$ pulses. The receiver combines the energy of those pulses subject to the bit-wise hypothesis and the XOR-mask and applies a binary hypothesis-test per bit.
The outcome is a received bit sequence $\rxmsg = (\rxbit{1}, \ldots ,\rxbit{\nbit})$.

\subsubsection*{Signal residual extraction}
\begin{figure*}[t]
	\begin{center}
		\includegraphics[scale=0.5]{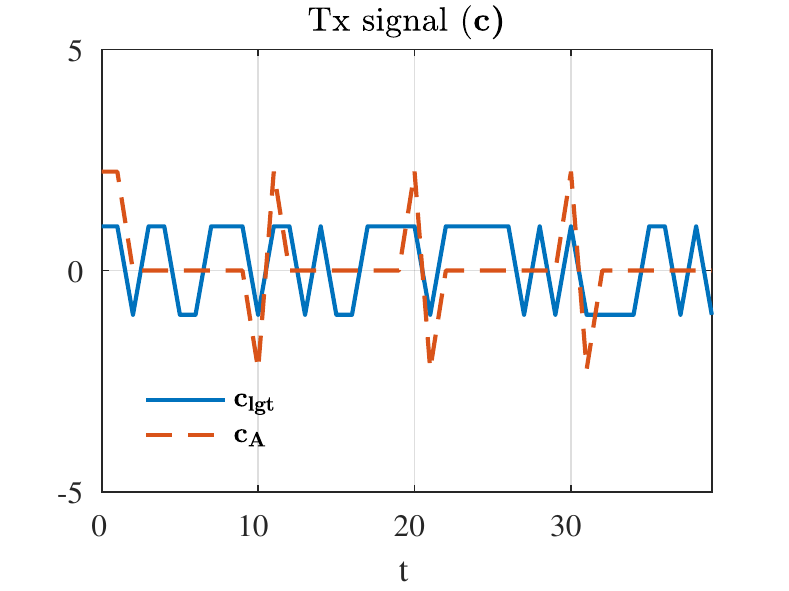}
		\includegraphics[scale=0.5]{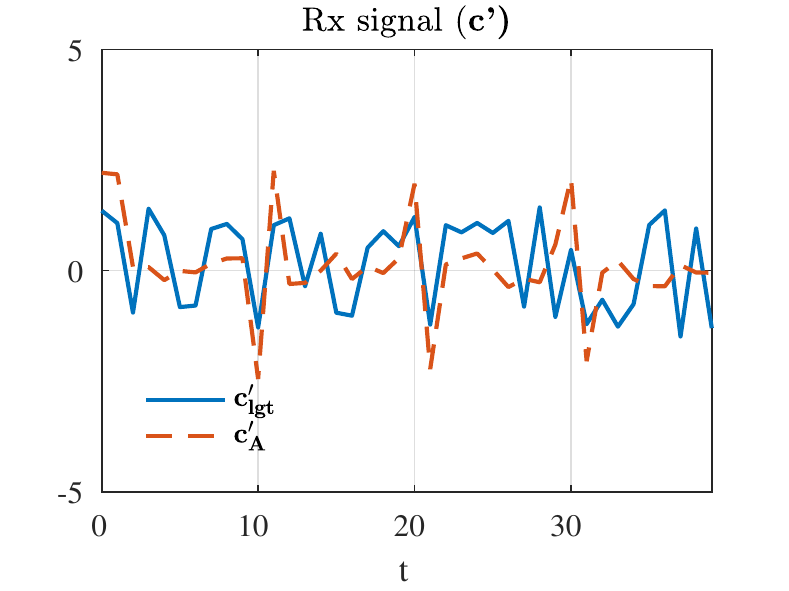}
		\includegraphics[scale=0.5]{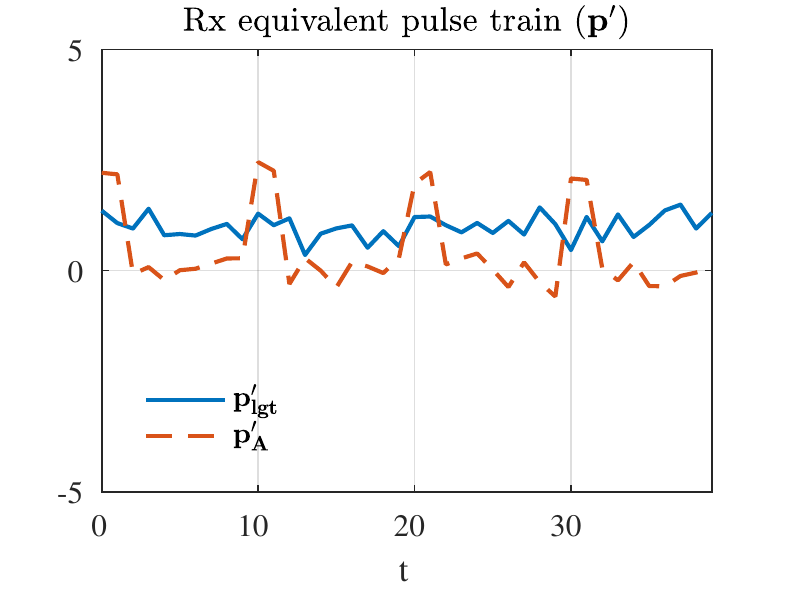}
		\includegraphics[scale=0.5]{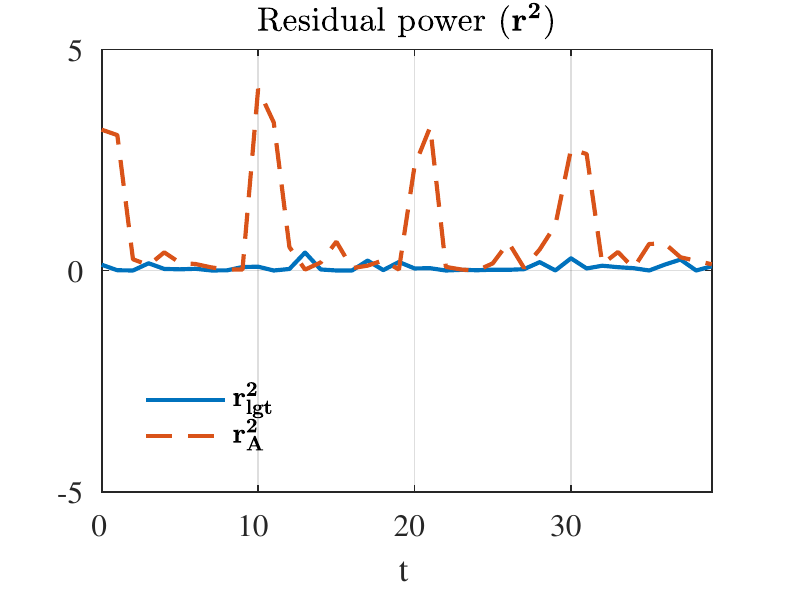}
		\caption{Legitimate (blue) and attack (red) signals in a scenario with four bits and 10 bits per symbol and repetition coding. The first plot shows the shape of the transmitted signal. The attack signal is winning since each bit contains sufficient power, despite the attacker only guessing 2 out of 10 pulses per symbol. The second plot shows the noisy signals at the receiver. The third plot shows the received signal after removing the data modulation. The residual after the expected signal component has been removed is shown in the rightmost plot. It becomes evident that the attack residual can be discerned easily from the legitimate residual, despite the attack on repetition coding (i.e., the bit level) being successful.}
		\label{fig:var_motivation}
	\end{center}
\end{figure*}

In order to test the signal integrity on the physical layer, we need to extract the signal-level residual. We exemplify the residual extraction at the receiver in Figure~\ref{fig:var_motivation}.
Under our stated assumptions about channel and modulation, the received signal $\rxsig$ consists of the actual signal $\txsig$, attenuated by path loss, as well as additive white Gaussian noise (AWGN).
At the receiver, the expected pulse polarity sequence (i.e., the template) $\rxtmpl$ is constructed based on the detected bits $\rxbit{i}$ and the shared XOR-sequence $\xorseq$, as in
$\rxtmpl = (\symb{\rxbit{1}} \Vert \ldots \Vert \symb{\rxbit{\nbit}}) \oplus \xorseq .$
We refer to this step as \emph{template generation} in Figure~\ref{fig:pipeline}.
The receiver-side equivalent pulse train is then given by the element-wise multiplication of the received signal with the expected pulse polarity sequence $\rxtmpl$, as $\mathbf{p}' = \rxsig \odot \rxtmpl$.
The residual is then obtained by subtracting the expected value from the receiver-side equivalent pulse train, as in $\ressig = \rxptr - \mu_{\rxptr}$.
A variance-based hypothesis test is concerned with whether the receive signal error consists of model error only or also contains an attacker error.
As we argue in Section~\ref{sec:design_space}, the property to test for is the variance of the signal residual, i.e., if the variance matches the expected noise or is too large, i.e., was caused by attacker errors. However, we require some normalization since the overall receive SNR will vary.

\subsubsection*{Secure distortion test}
We need to normalize the observed signal error to the overall signal energy. This way, we do not need to maintain an explicit noise estimate. The worst-case SNR is found by maximizing over the performance region $\perfreg$, guiding the choice of a threshold for the legitimate distortion. We are then able to check if the observed distortion, i.e., the overall normalized signal error, is within this bound.

This involves a hypothesis test on the normalized variance of the received signal, after being XOR-ed with the expected sequence. As secure distortion function, we propose taking the ratio between the power of the signal residual and the overall received power:
\[\dstrt = \frac{\ressig ^2}{\Vert \rxsig \Vert^2} = \frac{\sigma_{\rxptr}^2}{\Vert \rxsig \Vert^2} = \frac{\sum_i \left(\rxsig[i]\rxtmpl[i] - \frac{\sum_j \rxsig[j]\rxtmpl[j]}{\frmlen}\right)^2}{{\Vert \rxsig \Vert}^2}\]
The distortion can be interpreted as the inverse of a receive SNR estimate, based on a hypothesis on the pulse-level structure of the received signal. A random, zero-mean process will, for instance, evaluate to a distortion of $\dstrt = 1$.

Consequently, we can write the hypothesis test given by $\vrfyalgname$ as a decision between a signal containing some of the expected structure
\[\nullhypo :~ \dstrt < 1,\]
and the signal being only (attacker-induced) random noise:
\[\althypo :~ \dstrt = 1.\]

\subsubsection*{Performance region, decision threshold}
We assume the transmitter to choose the number of pulses per bit appropriately given a previously selected performance level $\perflvl = (\dist ',\ber ',\pl_{nlos}'),~\perflvl \in \perfreg$, i.e., such that
\[Q\left(\sqrt{\frac{\nppb \prx (\dist ',\pl_{nlos}')}{\sigma_n^2}}\right) \stackrel{!}{\leq} \ber '.\]

To satisfy our robustness criterion, the maximum legitimate signal distortion needs to be chosen such that the false negative rate does not exceed the underlying frame error rate, i.e.,
\[\thrs{\dstrt}(\perflvl) = \max{(\thrs{\dstrt}' \in [0, 1])}\text{~~,~s.t.~~}P[\dstrt_{lgt} > \thrs{\dstrt}'] \stackrel{!}{\leq} \fer.\]
The effective threshold is then chosen as the maximum threshold over the entire performance region, i.e., \mbox{$\thrseff{\dstrt} = \max_{\perflvl \in \perfreg} \thrs{\dstrt}(\perflvl)$}.
As a result, $\thrseff{\dstrt}$ results in a robust test under any performance tradeoff within the performance region $\perfreg$.

\subsection{\varmtac{}: Summary}
To summarize and illustrate how to embed the \varmtac{} into a distance-measurement system, we highlight the steps involved in the detection of an advancement attack by a receiver (\rx{}) on a signal originating from a transmitter (\tx{}).
\begin{mylist}
\item Pre-configuration
\begin{enumerate}
\item \rx{} determines the maximum accepted distortion threshold $\thrseff{\dstrt}$ based on the maximum communication distance and maximum tolerated noise level, subject to a performance region $\perfreg$.
\end{enumerate}
\item Key generation ($\genalgname$)
\begin{enumerate}
\item \tx{} and \rx{} derive a fresh pseudorandom XOR sequence $\xorseq$ from some shared secret. $\xorseq$ could theoretically also be secretly shared before each round\footnote{We don't have any requirements on ToA protection in this step.}.
\end{enumerate}
\item Mtac generation ($\mtacalgname$)
\begin{enumerate}
\item \tx{} encodes the message $\txmsg$ using repetition coding according to a chosen configuration $\perflvl \in \perfreg$ and applies the XOR sequence.
\end{enumerate}
\item Mtac verification ($\vrfyalgname$)
\begin{enumerate}
\item \rx{} constructs the message $\rxmsg$ by multiplying the received pulse sequence $\rxsig$ with the expected XOR sequence and applying a bit-wise binary hypothesis test on the overall symbol energy.
\item Based on the received message $\rxmsg$ and the XOR sequence, \rx{} constructs the expected pulse-level sequence $\rxtmpl$ (i.e., the template).
\item \rx{} computes the signal distortion $\dstrt (\rxsig, \rxtmpl)$ between received and expected pulse sequence.
\item \rx{} checks if $\dstrt$ exceeds $\thrseff{\dstrt}$. If so, it declares attack.
\end{enumerate}
\end{mylist}

\subsection{Practical concerns}
\subsubsection*{Time reference: Distance commitment}
We assume the detection of an advancement attack to be limited to verification of the data relative to some established time frame. This can be achieved by a distance commitment as introduced in~\cite{tippenhauer2015uwb}. This means the prover is assumed to have already responded in quick fashion to the query by transmitting a deterministic preamble, i.e., is committed to certain temporal reference. Relative to this temporal reference, the prover then has to deliver the secret information (i.e., $\txmsg$, correctly modulated) at a pre-agreed time relative to the preamble. It is realistic to assume a channel to be coherent throughout the frame, as the duration of a UWB frame used for distance measurement is typically less than 1ms.
Through a distance commitment, the vulnerabilities of a back-search~\cite{poturalski2010cicada} on the data-bearing part can be avoided.

\subsubsection*{Ranging precision}
Under a distance commitment, the back-search for the acquisition of the first signal path is only necessary on the preamble of the frame. Therefore, the precision of the ranging procedure is not determined by any operation applied to the data-bearing part. Consequently, the precision of our proposal cannot be worse than that of existing schemes relying on a distance commitment. It has been shown that such a system can achieve a precision of 10cm, irrespective of communication distance~\cite{deca,singhuwb}.

\subsubsection*{Bit-level security}
We assume a bit-level procedure to detect if the received bits $\rxmsg$ do not match the transmitted message $\txmsg$. This could be achieved by a message authentication code (MAC) appended to the frame or even transmitted on a separate, potentially ToA-agnostic channel.

%!TEX root =  mtac_main.tex
\section{Analysis}
\label{sec:analysis}
\newcommand{\numBits}{32}
\newcommand{\secDistLoS}{200m}
\newcommand{\secDistNLoS}{20m}

In the following, we explore the tradeoff between security and performance by modeling the effect of the channel and evaluating the classification performance of our \varmtac{} from the previous section. The results are based on simulations, which, however, make assumptions in line with realistic UWB-based distance measurement systems.
From these results, we can derive the performance region in which our proposal maintains bit-equivalent security (i.e., $\adv(\frgrnameworst)<2^{-\nbit}$) and how to scale to longer distances.

\subsection{Model}
\subsubsection*{Path loss model}
\begin{figure}
    \centering
    \begin{minipage}{0.24\textwidth}
        \centering
        \includegraphics[width=1\textwidth]{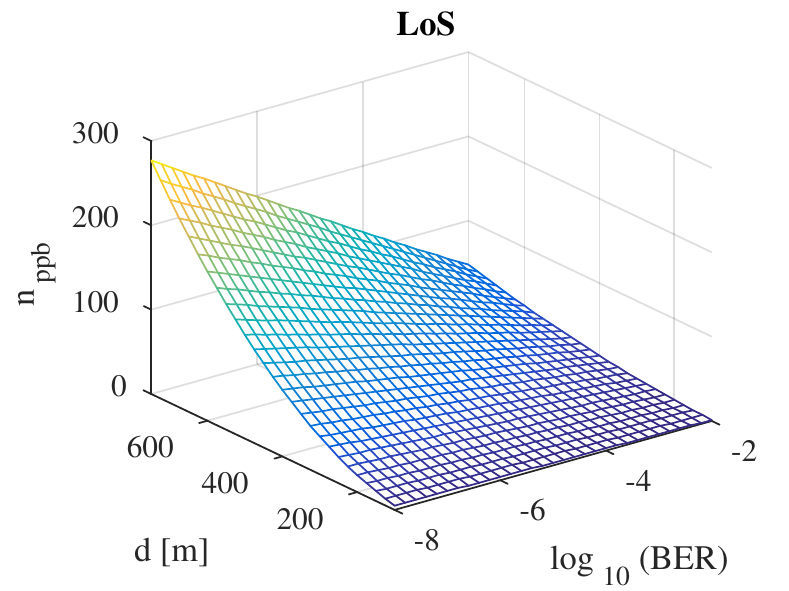}
    \end{minipage}\hfill
    \begin{minipage}{0.24\textwidth}
        \centering
        \includegraphics[width=1\textwidth]{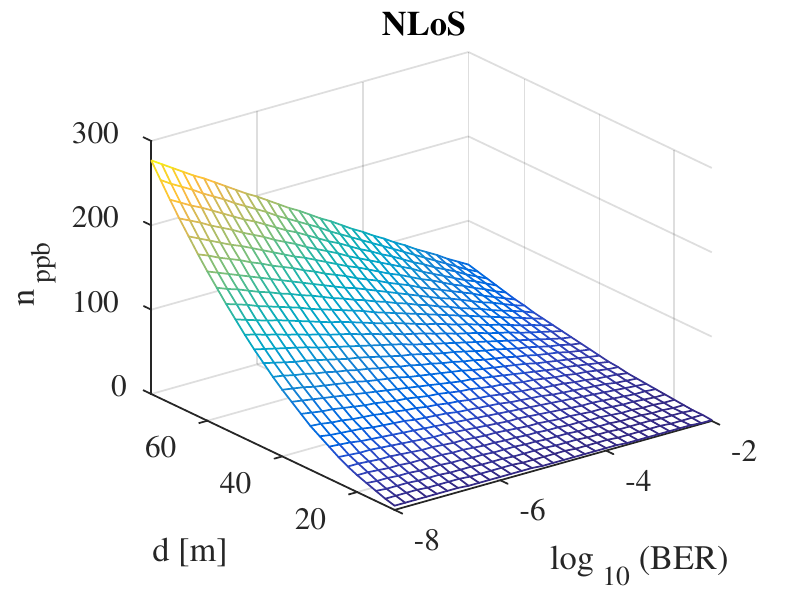}
    \end{minipage}
    \caption{The number of pulses per symbol as a function of the target performance level, i.e., the target bit error rate (BER) and operating distance under FCC/ETSI constraints. These numbers refer to a LoS (left) as well as a NLoS scenario (right) with 20dB attenuation of the direct path. Lower BERs over longer distances require more pulses per symbol.}
    \label{fig:n_ppb}
\end{figure}

To evaluate the impact of distance on a) the modulation required and b) the implications on security, we assume a free-space path loss model. This means the received power degrades inversely to the square of the distance, as in
\[\prx = \ptx \left(\frac{\lambda}{4\pi \dist}\right)^2 \pl_{nlos}.\]
%and the BER is given by Equation~\ref{eq:ber}.
We assume the antennas to be operated in each other's far field, as the goal of this analysis is to understand the tension between long distance and security.
As input power, we rely on the constraints put forward by the FCC and ETSI regarding UWB in licensed spectrum. This is, a maximum peak power of 0dBm within the 50MHz around the peak and an average limitation on signal power spectral density of {-41.3dBm/Hz}. We assume that our pulses are sufficiently spaced, such that each pulse can be sent at peak power. We assume a signal bandwidth of 620MHz at a center frequency of 6681.6MHz, which is a typical UWB channel configuration~\cite{4z_standard}. For receiver-side noise, we consider the thermal noise figure at room temperature, given by \hbox{-174dBm/Hz}. In a separate non-line-of-sight (NLoS) scenario, we assume an additional attenuation of 20dB which is roughly the attenuation the signal experiences when traversing the human body. In Figure~\ref{fig:n_ppb}, we show the number of pulses per symbol required under both LoS and NLoS conditions. The required number of pulses increases with longer distances and decreases if the requirement on target BER gets relaxed.

\subsubsection*{Gaussian model for variance distributions}
The variance constitutes a sum of $\frmlen$ independent random variables. Due to the central limit theorem, for a sufficiently high overall number of pulses, the variance distribution converges to a Gaussian, i.e.,
\begin{equation}
\label{eq:dist_gauss_att}
\dstrt_{\frgrnameworst}(\dist) \sim \nprc\left(\mu_{\dstrt_{\frgrnameworst}}(\dist),\sigma_{\dstrt_{\frgrnameworst}}(\dist)\right)
\end{equation}
\begin{equation}
\label{eq:dist_gauss_lgt}
\dstrt_{lgt}(\dist) \sim \nprc\left(\mu_{\dstrt_{lgt}}(\dist),\sigma_{\dstrt_{lgt}}(\dist)\right).
\end{equation}
In general, these distributions are a function of the communication range as well as the target BER. Through simulations, we can verify that in the area of interest (i.e., where the distributions significantly overlap), these distributions indeed fit a Gaussian hypothesis well, as we show in detail in Appendix~\ref{sec:app_validation}.

\subsection{Results}
\begin{figure}
    \centering
    \begin{minipage}{0.24\textwidth}
        \centering
        \includegraphics[width=1\textwidth]{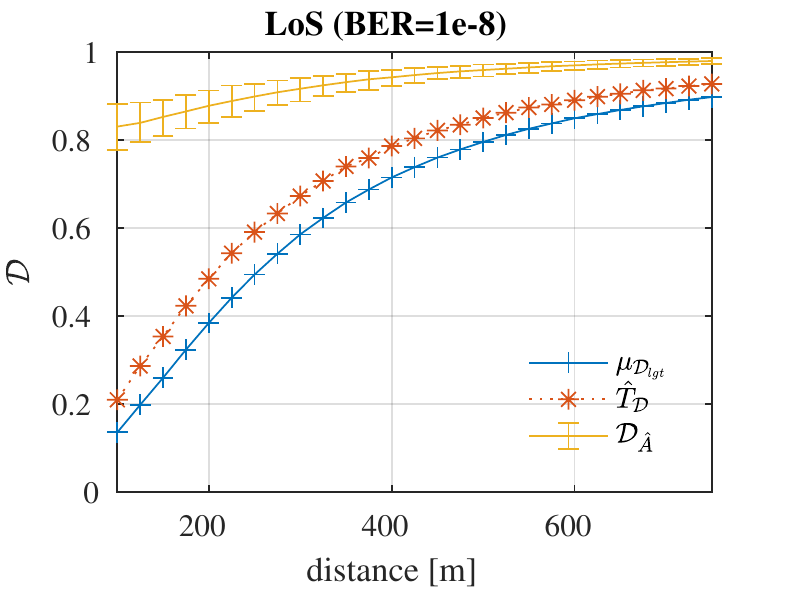}
    \end{minipage}\hfill
    \begin{minipage}{0.24\textwidth}
        \centering
        \includegraphics[width=1\textwidth]{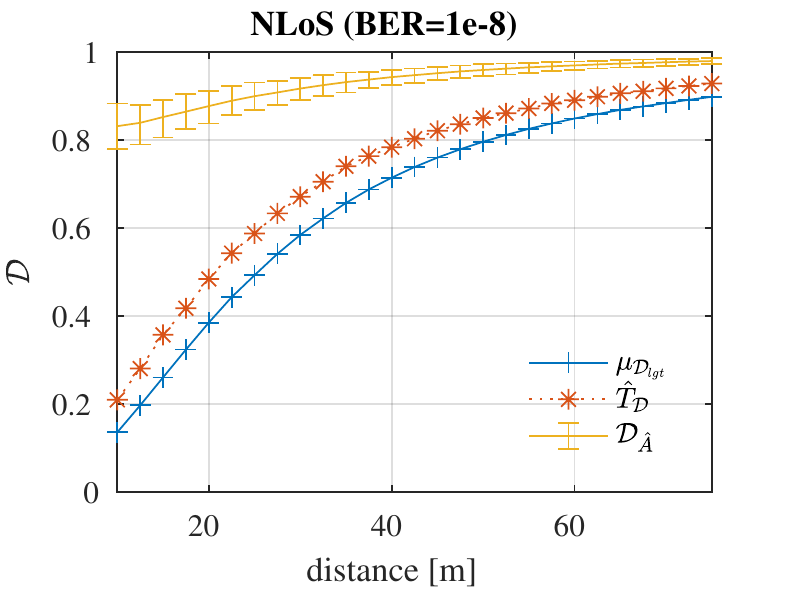}
    \end{minipage}
    \caption{Over longer distances, the legitimate distortion increases. The gap between maximum legitimate distortion and minimum attack distortion becomes smaller for longer distances, eventually vanishing altogether. This means, under our strong attacker model, MTAC security can only be maintained up to some distance.}
    \label{fig:stats_vs_dist}
\end{figure}

We model the bit error rate of the underlying modulation according to Equation~\ref{eq:ber}. We simulate this in MATLAB for a frame of \numBits{} bits. As we detail in Appendix~\ref{sec:app_frame_len}, the security guarantees are maintained for longer frames. For robustness, the choice of the decision threshold should result in the same false negative rate of $\vrfyalgname$ as under bit-wise detection, i.e., $\fnr{\vrfyalgname}\stackrel{!}{=}1-(1-\ber)^{\nbit}$. Under the Gaussian hypothesis for the distortion distribution, we can derive the practical decision threshold by choosing it $Q^{-1}(\fnr{\vrfyalgname})$ normalized standard deviations above the expected legitimate distortion. The resulting threshold is indicated in Figure~\ref{fig:stats_vs_dist}. We evaluate the probability of attacker success for a given maximum communication distance based on the attacker's best case statistics and the legitimate worst-case statistics, over a range of target BER values. This is in line with our attacker model, which does not make any assumptions about the attacker's position.
For a given performance region, the upper bound of the attacker's advantage is given by

\[\adv(\frgrnameworst) = Q\left(\frac{\hat{\mu}_{\dstrt_{\frgrnameworst}} - (\hat{\mu}_{\dstrt_{lgt}} + Q^{-1}(\fnr{\vrfyalgname})\cdot\hat{\sigma}_{\dstrt_{lgt}})}{\hat{\sigma}_{\dstrt_{\frgrnameworst}}}\right),\]

whereas the statistical parameters, i.e., means and variances, are chosen in favor of forger $\frgrnameworst$.
Specifically, we choose the attacker's parameters under minimization of the worst-case distortion
%, i.e.,
%\[(\hat{\mu}_{\dstrt_{att}}, \hat{\sigma}_{\dstrt_{att}}) = (\mu_{\dstrt_{att}}(d_{att,ideal}), \sigma_{\dstrt_{att}}(d_{att,ideal}))\]
%\[d_{att,ideal} = \argmin_{d\in[0, d_{max}]}{\mu_{\dstrt_{att}}}(d)-\sigma_{\dstrt_{att}}(d),\]
and the parameters of the legitimate transmitter under maximization of the distortion, within the defined performance region. The details of those choices we provide in Appendix~\ref{sec:app_stat_param}.
%i.e.,
%\[(\hat{\mu}_{\dstrt_{lgt}}, \hat{\sigma}_{\dstrt_{lgt}}) = (\mu_{\dstrt_{lgt}}(d_{lgt,worst}), \sigma_{\dstrt_{lgt}}(d_{lgt,worst}))\]
%\[d_{lgt,worst} = \argmax_{d\in[0, d_{max}]}{\mu_{\dstrt_{lgt}}}(d)+\sigma_{\dstrt_{lgt}}(d).\]
Unsurprisingly, the worst-case distance for the legitimate transmitter amounts typically to the maximum distance.
%, i.e., $d_{lgt,worst}=d_{max}$.
The numerical values of those statistical parameters (i.e., means and variances) were obtained through simulation. We thereby modeled the attacker as having an ideal pulse-level bias of $20\%$, as motivated in Section~\ref{sec:design_space}. In the following, we are interested in the performance region in which the MTAC provides bit-equivalent security, i.e., $\adv(\frgrnameworst)\leq 2^{-\nbit}$.

\subsubsection*{Performance-equivalent MTAC region}
\begin{figure}
    \centering
    \begin{minipage}{0.24\textwidth}
        \centering
        \includegraphics[width=1\textwidth]{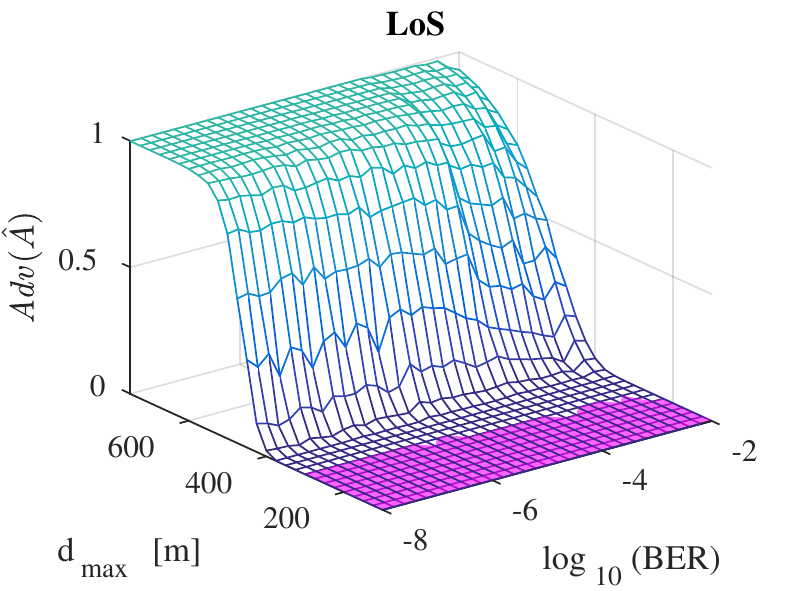}
    \end{minipage}\hfill
    \begin{minipage}{0.24\textwidth}
        \centering
        \includegraphics[width=1\textwidth]{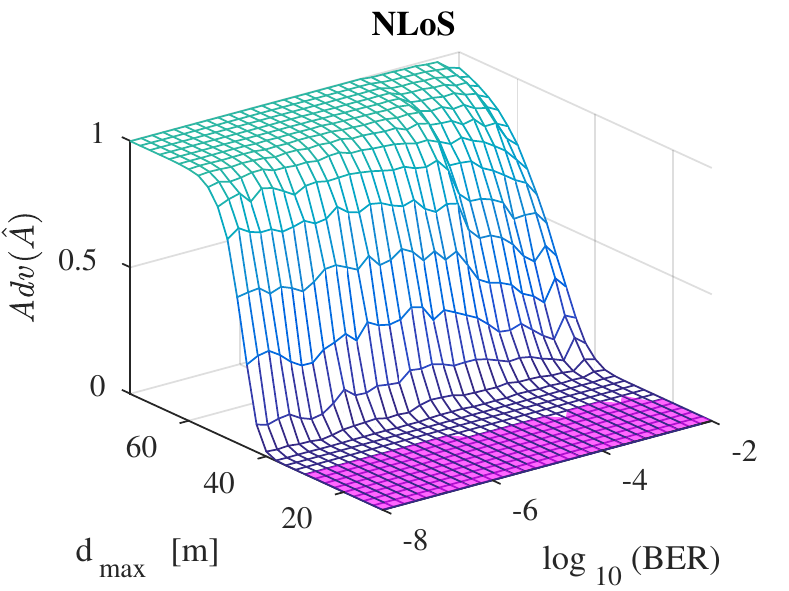}
    \end{minipage}
    \caption{Attacker's advantage as a function of the performance level. We highlight the performance region within which the MTAC provides bit-equivalent security. The secure distortion test provides us with a bit-equivalently secure MTAC for distances up to \secDistLoS{} and \secDistNLoS{} for LoS and NLoS scenarios, respectively.}
    \label{fig:region_equiv}
\end{figure}

Figure~\ref{fig:region_equiv} shows the attacker's advantage as a function of the performance level. The figure highlights the performance region in which we have bit-equivalent MTAC security.
%Under LoS conditions, we have an MTAC up to a range of roughly 1000m. Under NLoS conditions, this figure degrades to around 100m.
\begin{observation}
Under any tradeoff between symbol length and target bit error rate: For any frame $\txmsg$ of at least \numBits{} bits, we can find a distortion threshold $\thrseff{\dstrt}$ resulting in an MTAC with bit-equivalent security for distances up to \secDistLoS{} under LoS conditions and up to \secDistNLoS{} under NLoS conditions.
\end{observation}

\subsubsection*{Extending the MTAC region}
By comparing the results for LoS and NLoS conditions, we see that the MTAC region seems to degrade proportionally to the attenuation added, i.e., the results are invariant under amplification/attenuation. This means we can extrapolate to any communication range if we allocate a security link margin $\pl_{sec}\geq 0$ satisfying
\[\pl_{sec} \stackrel{!}{\geq} 20\cdot \log_{10}\left(\frac{\dist_{max}}{\secDistLoS{}}\right) + \pl_{nlos}.\]

%!TEX root =  mtac_main.tex
\section{Conclusion}
\label{sec:conclusion}
With MTAC, we propose a physical-layer primitive for secure distance measurement. We formally define the security of its underlying algorithms. We then derive design principles for the practical instantiation of an MTAC: A randomized pulse sequence and a secure distortion test over the entire signal.
The results indicate that the bit-equivalent security level can be regained over a meaningful performance region, thereby resulting in a fundamental building block preventing any physical-layer, distance-reducing attacks.

\section{Acknowledgements}
This project has received funding from the European Research Council (ERC) under the European Union's Horizon 2020 research and innovation programme under grant agreement No 726227.

\newpage

\bibliographystyle{IEEEtran}
%\balance
\bibliography{mtac}

\begin{appendices}
%!TEX root =  mtac_main.tex
\section{Lessons Learned}
\label{sec:lessons_learned}
Since our proposal provides a fundamentally secure physical-layer building block for ToA measurement, this changes prevailing assumptions on the design of higher level protocols. In~\cite{clulow2006so}, Clulow et al. put forward principles for distance bounding:
\blockquote{"We propose a number of principles to adhere to when implementing distance-bounding systems. These restrict the choice of the communication medium to speed-of-light channels, the communication format to single bit exchanges for timing, symbol length to narrow (ultra wideband) pulses, and protocols to error-tolerant versions. These restrictions increase the technical challenge of implementing secure distance bounding."}

Given the above designs, these recommendations don't seem to hold up. 
Also, the recommendation to single bit timing is not only not needed but also are fairly wasteful.  Again, from~\cite{clulow2006so}:

\blockquote{"We show that proposed distance-bounding protocols of Hu, Perrig, and Johnson (2003), Sastry, Shankar and Wagner (2003), and {\v{C}}apkun and Hubaux (2005, 2006) are vulnerable to a guessing attack where the malicious prover preemptively transmits guessed values for a number of response bits."}

In this work, we show that these vulnerabilities are an artifact of a somewhat naively designed physical layer and modulation, and can be addressed purely on the physical layer. The problem and its solution are orthogonal to the design and security of the protocol that builds in it, which operates on a different level of abstraction, making different assumptions about the security parameter (i.e., the bits) making up the nonces. These protocols not addressing the physical layer does not mean they are necessarily vulnerable. If coupled with a physical layer that is in line with our design, they are secure, within the performance region we point out.

%!TEX root =  mtac_main.tex
\section{Attacker Statistical Parameters}
\label{sec:app_stat_param}
We choose the attacker's parameters under minimization of the worst-case distortion, i.e., as
\[(\hat{\mu}_{\dstrt_{\frgrnameworst}}, \hat{\sigma}_{\dstrt_{\frgrnameworst}}) = (\mu_{\dstrt_{\frgrnameworst}}(\dist_{\frgrnameworst,ideal}), \sigma_{\dstrt_{\frgrnameworst}}(\dist_{\frgrnameworst,ideal}))\]
\[\dist_{\frgrnameworst,ideal} = \argmin_{\dist\in[0, \dist_{max}]}{\mu_{\dstrt_{\frgrnameworst}}}(\dist)-\sigma_{\dstrt_{\frgrnameworst}}(\dist),\]
and the parameters of the legitimate transmitter under maximization of the distortion, within the defined performance region, i.e., as
\[(\hat{\mu}_{\dstrt_{lgt}}, \hat{\sigma}_{\dstrt_{lgt}}) = (\mu_{\dstrt_{lgt}}(\dist_{lgt,worst}), \sigma_{\dstrt_{lgt}}(\dist_{lgt,worst}))\]
\[\dist_{lgt,worst} = \argmax_{\dist\in[0, \dist_{max}]}{\mu_{\dstrt_{lgt}}}(\dist)+\sigma_{\dstrt_{lgt}}(\dist).\]

%!TEX root =  mtac_main.tex
\section{Validating the Gaussian Variance Model}
\label{sec:app_validation}
\newcommand{\numBitsQQ}{32}
\newcommand{\numBitsEmp}{20}
\newcommand{\attRlvtDistLoS}{100m}
\newcommand{\attRlvtDistNLoS}{10m}
\newcommand{\lgtRlvtDistLoS}{200m}
\newcommand{\lgtRlvtDistNLoS}{20m}

In the following, we motivate the Gaussian model for the distortion distribution put forward in Equations~\ref{eq:dist_gauss_att}~and~\ref{eq:dist_gauss_lgt}.

\subsection{Extrapolation vs. fully empirical results}
\begin{figure}
	\centering
	\begin{minipage}{0.24\textwidth}
		\centering
		\includegraphics[width=1\textwidth]{plots/pwin_actual_los.pdf}
	\end{minipage}\hfill
	\begin{minipage}{0.24\textwidth}
		\centering
		\includegraphics[width=1\textwidth]{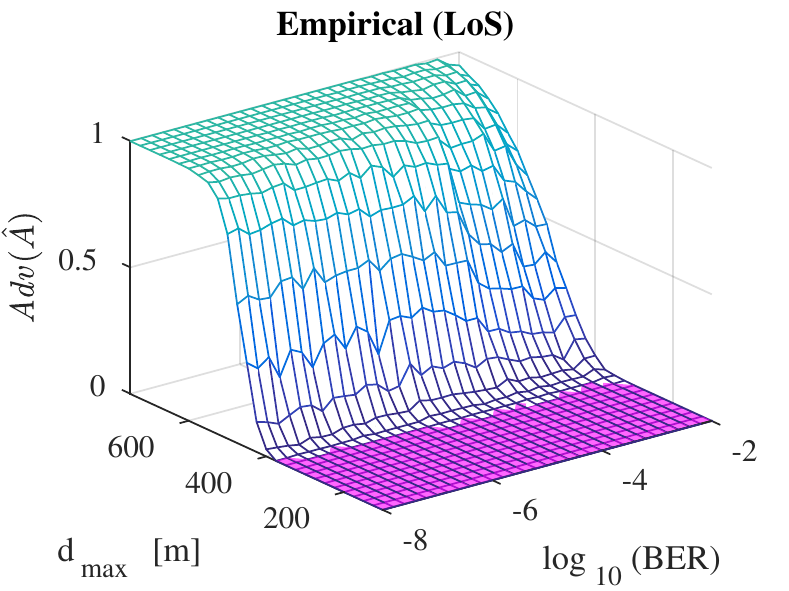}
	\end{minipage}
	\caption{Attacker's advantage as a function of the performance level for LoS conditions under a Gaussian extrapolation (left) and fully empirical simulation (right). Overall, the empirical result is very similar, in particular its MTAC region is not smaller than the one resulting from the Gaussian model.}
	\label{fig:pwin_vs_emp_los}
\end{figure}

\begin{figure}
	\centering
	\begin{minipage}{0.24\textwidth}
		\centering
		\includegraphics[width=1\textwidth]{plots/pwin_actual_nlos.pdf}
	\end{minipage}\hfill
	\begin{minipage}{0.24\textwidth}
		\centering
		\includegraphics[width=1\textwidth]{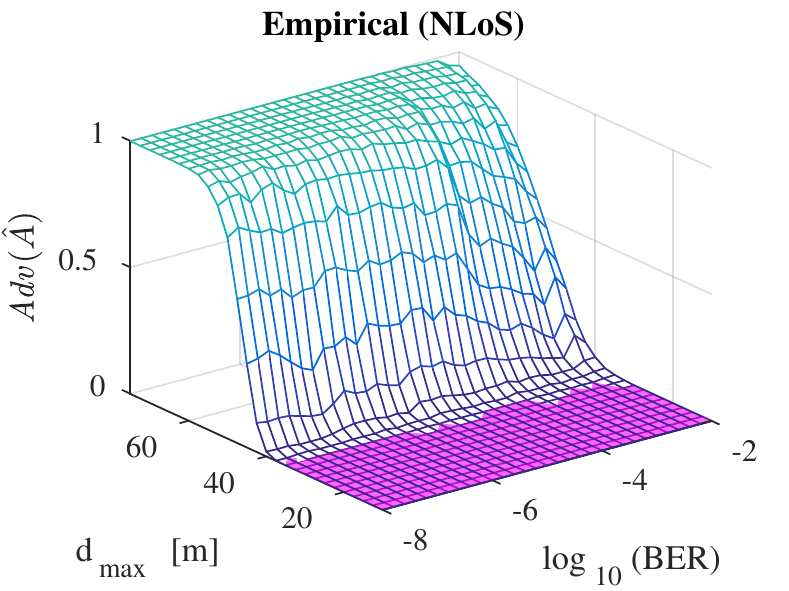}
	\end{minipage}
	\caption{Attacker's advantage as a function of the performance level for NLoS conditions under a Gaussian extrapolation (left) and fully empirical simulation (right). Overall, the empirical result is very similar, in particular its MTAC region is not smaller than the one resulting from the Gaussian model.}
	\label{fig:pwin_vs_emp_nlos}
\end{figure}

In the following, we compare our extrapolated results from Section~\ref{sec:analysis} to a fully empirical (i.e., Monte-Carlo) simulation. The probability of winning as a function of the performance level is shown in Figure~\ref{fig:pwin_vs_emp_los} for LoS conditions and Figure~\ref{fig:pwin_vs_emp_nlos} for NLoS conditions.
Both results refer to a frame of \numBitsEmp{} bits. For both scenarios, we see that the the attacker's advantage evolves almost identically. We see that the fully empirical results indicate a slightly wider MTAC region, which suggest our Gaussian model to be a conservative estimate.

\subsection{Variance distribution is sufficiently Gaussian}
\begin{figure*}[t]
	\begin{center}
		\includegraphics[scale=0.5]{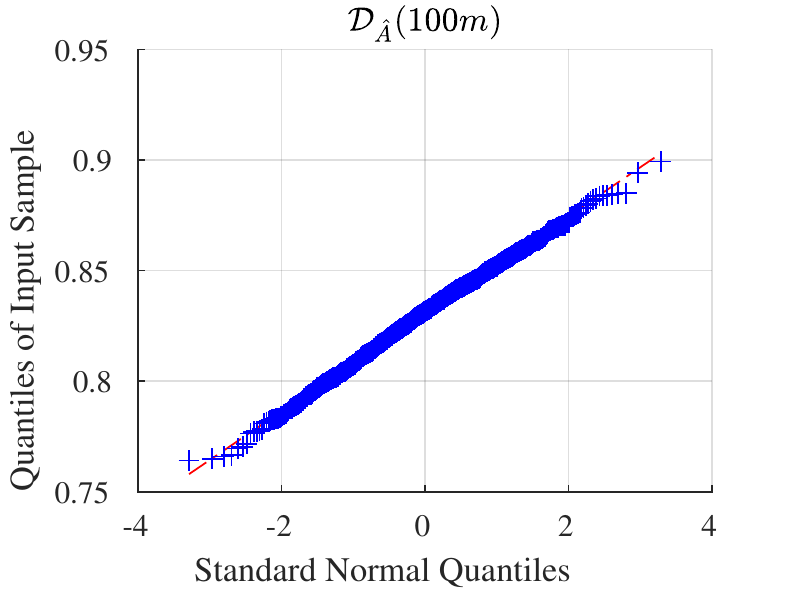}
		\includegraphics[scale=0.5]{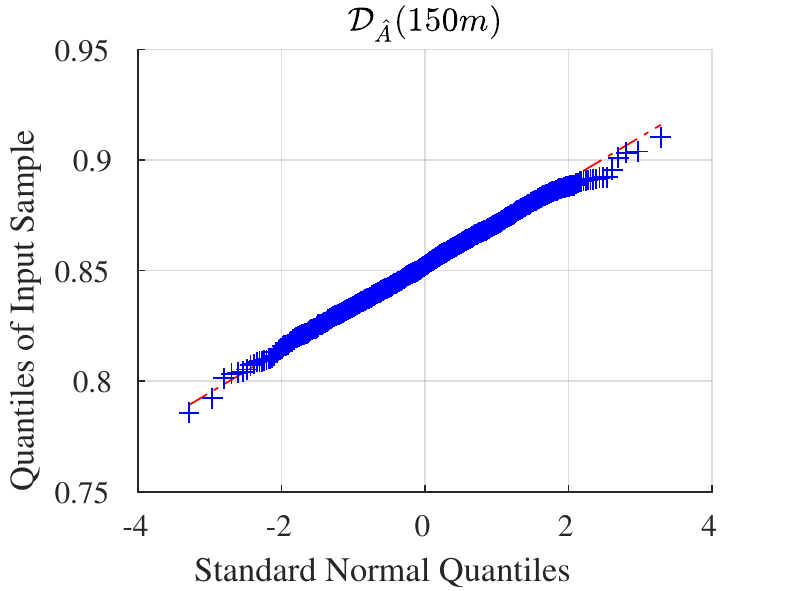}
		\includegraphics[scale=0.5]{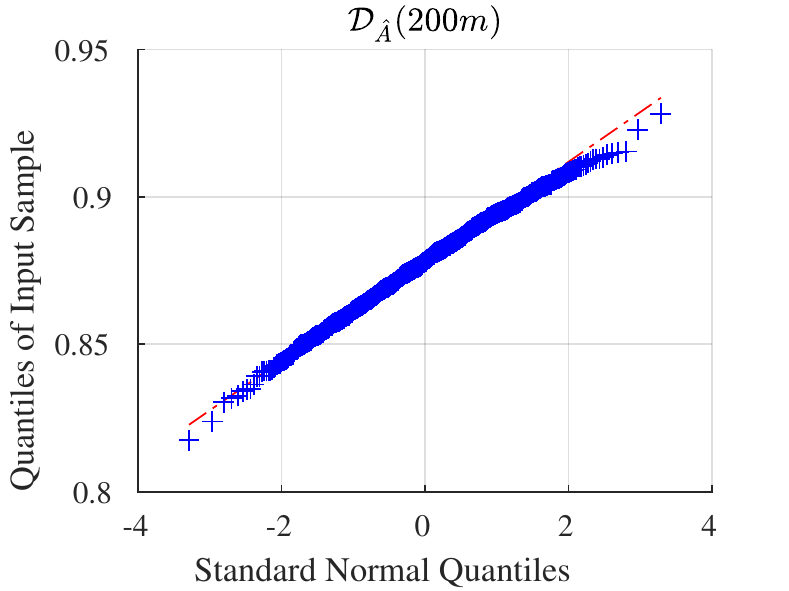}
		\includegraphics[scale=0.5]{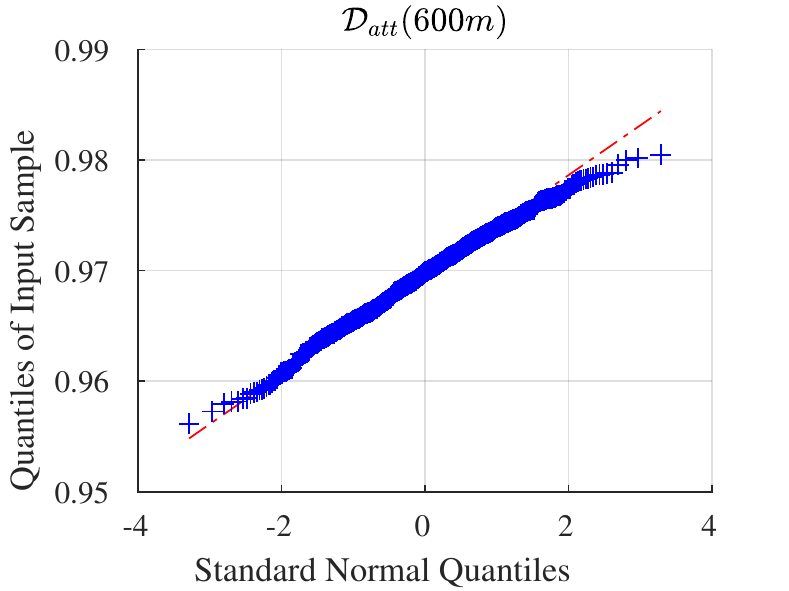}
		\includegraphics[scale=0.5]{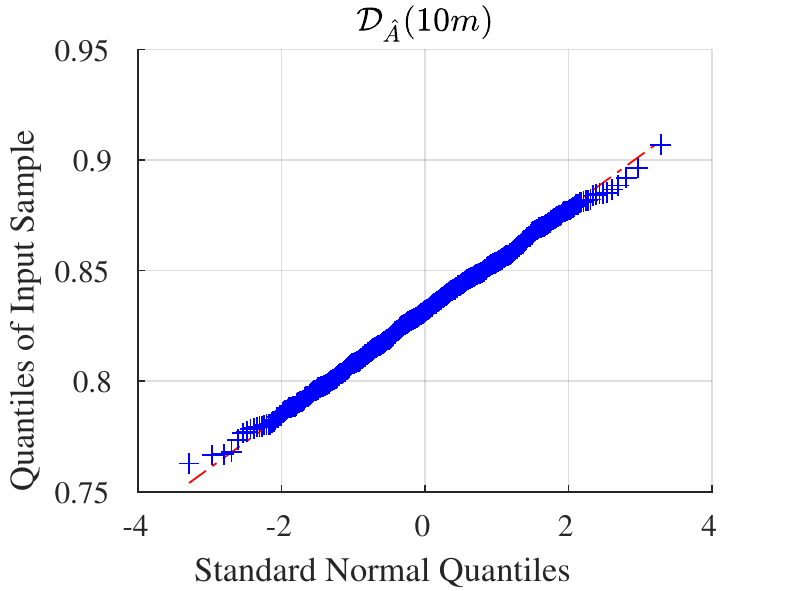}
		\includegraphics[scale=0.5]{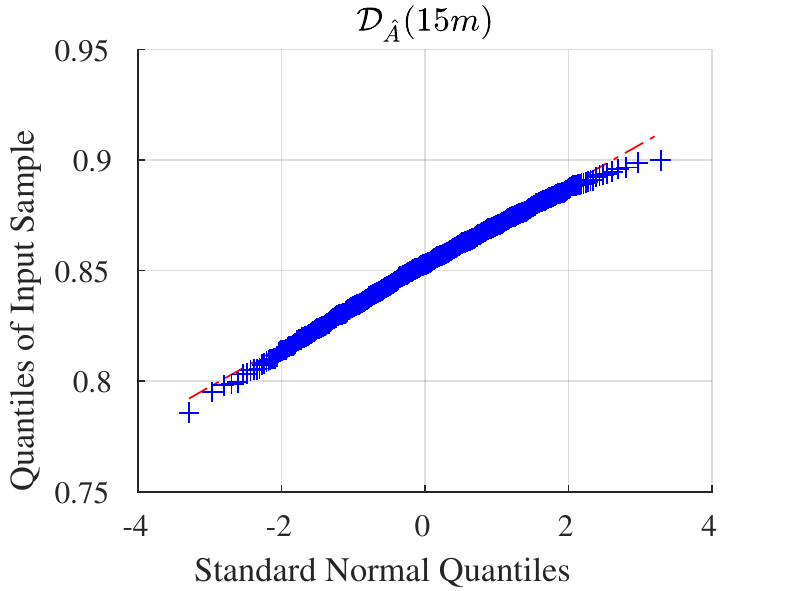}
		\includegraphics[scale=0.5]{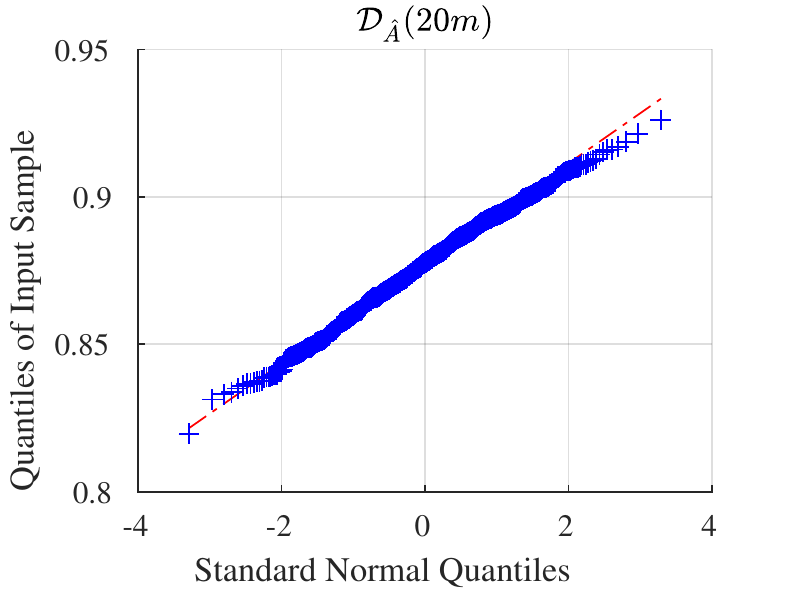}
		\includegraphics[scale=0.5]{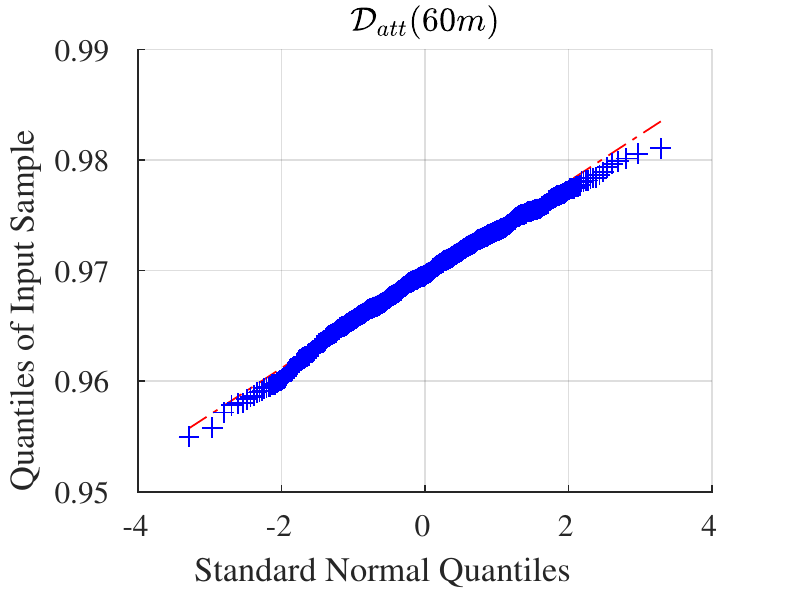}
		\caption{QQ plots comparing the attacker empirical distortion distribution for LoS (top) and NLoS (bottom) conditions for a frame of \numBitsQQ{} bits and different distances against a normal distribution. For validity of results w.r.t. the MTAC region boundary, the attack signal distortion at a distance of \attRlvtDistLoS{} (LoS) and \attRlvtDistNLoS{} (NLoS) should be close to a Gaussian. Indeed, the QQ plots of the second column are close to the diagonal.}
		\label{fig:qq_att}
	\end{center}
\end{figure*}

\begin{figure*}[t]
	\begin{center}
		\includegraphics[scale=0.5]{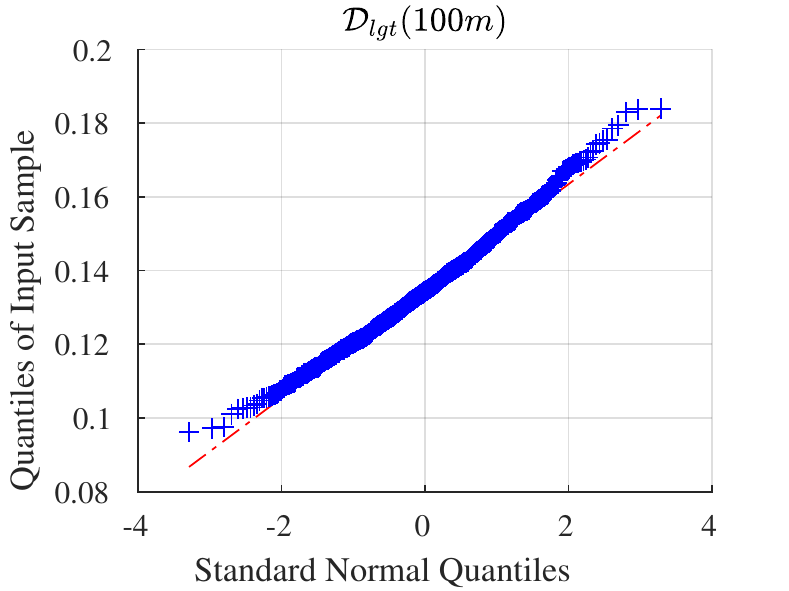}
		\includegraphics[scale=0.5]{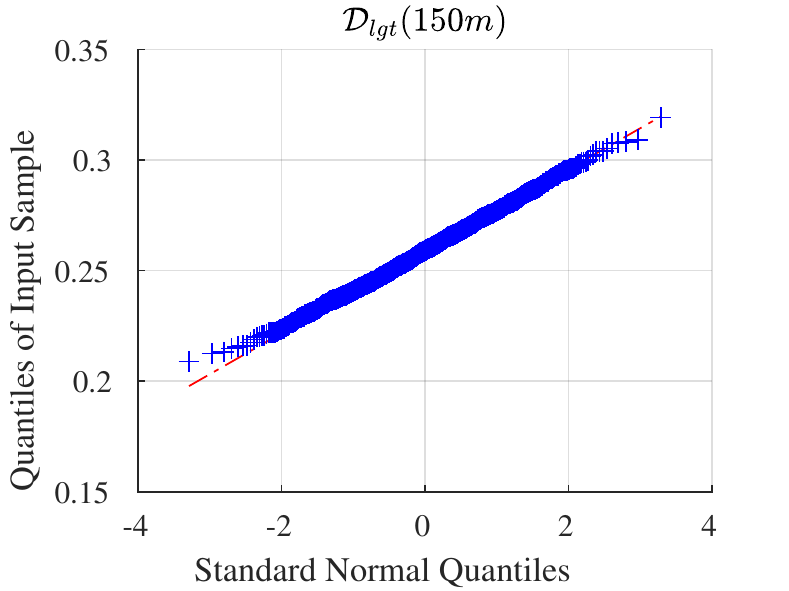}
		\includegraphics[scale=0.5]{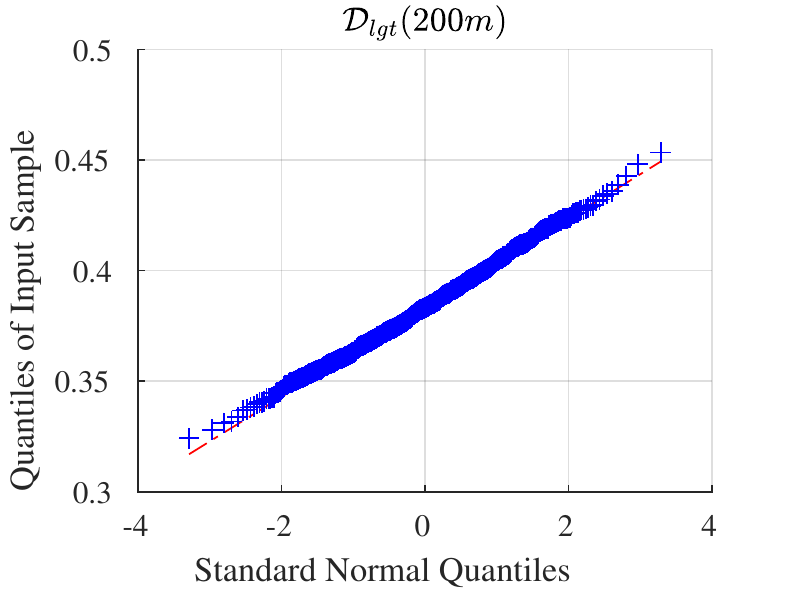}
		\includegraphics[scale=0.5]{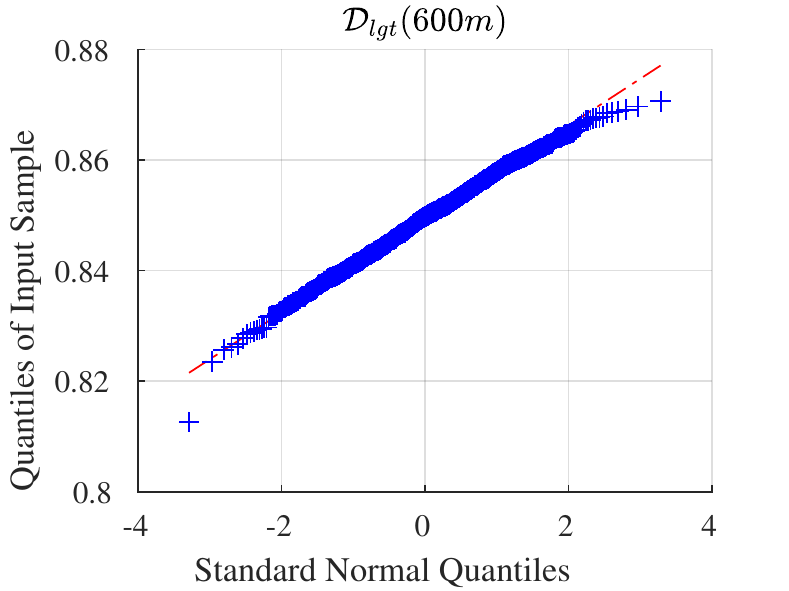}
		\includegraphics[scale=0.5]{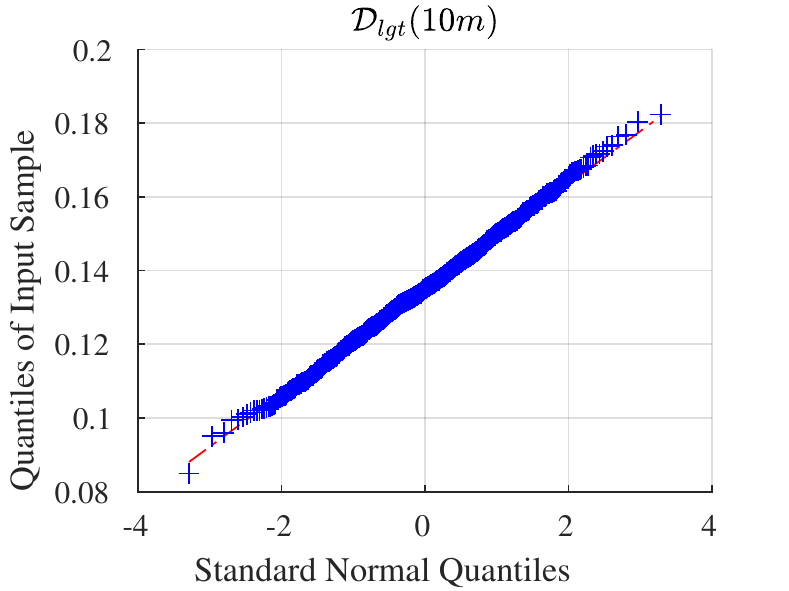}
		\includegraphics[scale=0.5]{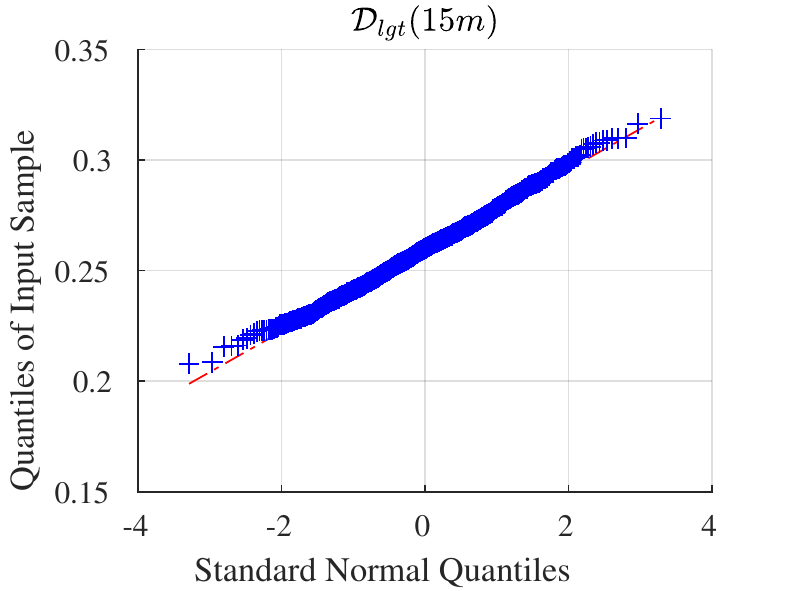}
		\includegraphics[scale=0.5]{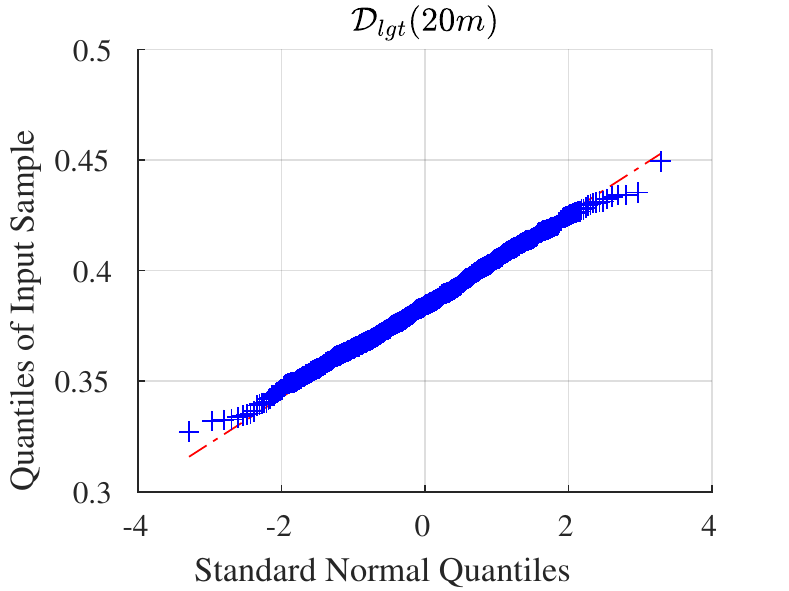}
		\includegraphics[scale=0.5]{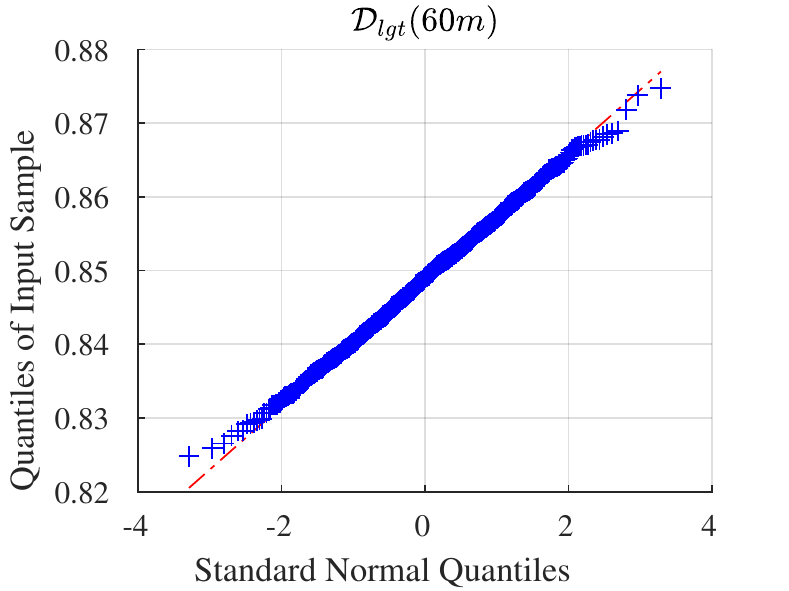}
		\caption{QQ plots comparing the legitimate empirical distortion distribution for LoS (top) and NLoS (bottom) conditions for a frame of \numBitsQQ{} bits and different distances against a normal distribution. For validity of results w.r.t. the MTAC region boundary, the attack signal distortion at a distance of \lgtRlvtDistLoS{} (LoS) and \lgtRlvtDistNLoS{} (NLoS) should be close to a Gaussian. Indeed, the QQ plots of the third column are close to the diagonal.}
		\label{fig:qq_lgt}
	\end{center}
\end{figure*}

We provide quantile-quantile (QQ) plots that compare the empirical distributions against normal distributions. This allows to validate the model we use in Section~\ref{sec:analysis} which serves extrapolate the empirical classification performance to small likelihoods. We provide those plots for a frame of \numBitsQQ{} bits and a selection of communication distances, both for LoS and NLoS scenarios. Figure~\ref{fig:qq_att} presents those results for the attacker's variance distribution. The relevant distance for the resulting MTAC region boundary is around \attRlvtDistLoS{} for LoS and around \attRlvtDistNLoS{} for NLoS. This is the distance at which the distortion for the attacker is minimal, see Figure~\ref{fig:stats_vs_dist}. There is a slight downwards bend of the empirical value for higher quantiles. This means, a slightly bit more than expected high-variance outliers compared to the Gaussian hypothesis. This is in line with our requirements, i.e., the normal estimate being conservative regarding distinguishability. The plots for those distances show that the empirical quantiles are well aligned with the straight diagonal.
Figure~\ref{fig:qq_lgt} presents those results for the attacker's distortion distribution. The relevant distance for the resulting MTAC boundary is around \lgtRlvtDistLoS{} for LoS and around \lgtRlvtDistNLoS{} for NLoS, i.e., mid-range. The plots for those distances show that the empirical quantiles are well aligned with the straight line at those distances relevant for the MTAC region derived in Section~\ref{sec:analysis}.

%!TEX root =  mtac_main.tex
\section{Effect of Frame Length}
\label{sec:app_frame_len}
\begin{figure}
    \centering
    \begin{minipage}{0.24\textwidth}
        \centering
        \includegraphics[width=1\textwidth]{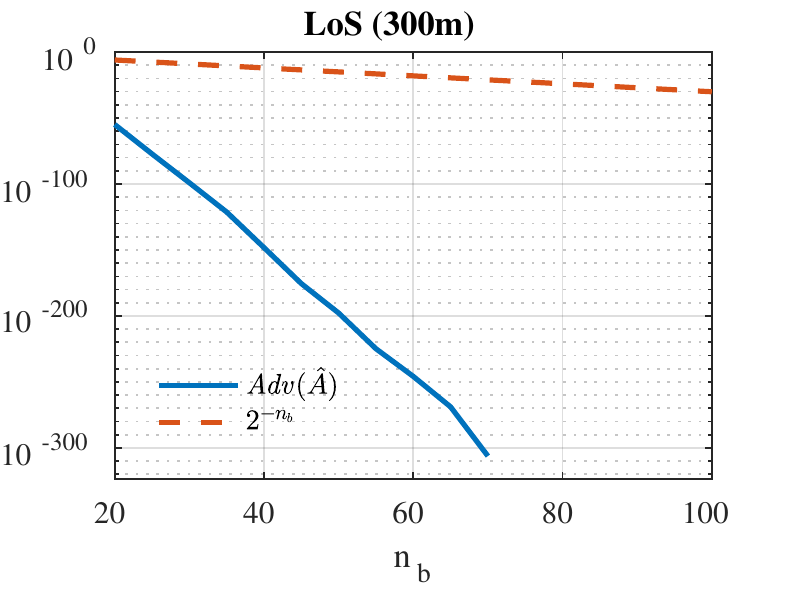}
    \end{minipage}\hfill
    \begin{minipage}{0.24\textwidth}
        \centering
        \includegraphics[width=1\textwidth]{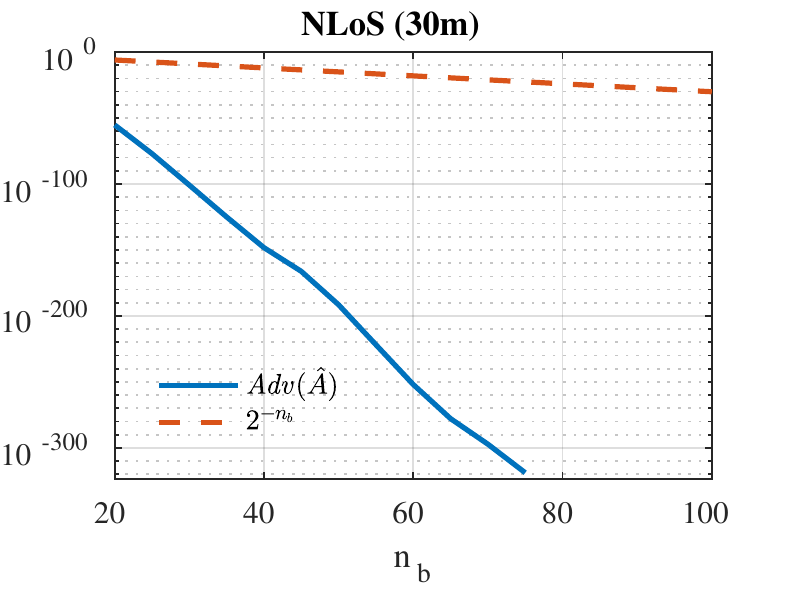}
    \end{minipage}
    \caption{The attacker's advantage decreases faster than the bit-equivalent MTAC security level for longer frames. This means the security guarantees for shorter frames are maintained for longer frames.}
    \label{fig:dist_vs_nbit}
\end{figure}

Figure~\ref{fig:dist_vs_nbit} shows the security level for one particular distance as a function of the frame length. It becomes evident that bit-level equivalence of the security level is maintianed as the length of the frame increases. We see that attacker's advantage decays faster than $2^{-n_b}$.

\end{appendices}

\end{document}